\documentclass[preprint]{JHEP3} 

\usepackage{graphicx}
\usepackage{amsmath}
\usepackage{comment}

\newcommand{\cA}{{\cal A}}
\newcommand{\cD}{{\cal D}}
\newcommand{\cF}{{\cal F}}
\newcommand{\cN}{{\cal N}}
\newcommand{\cQ}{{\cal Q}}
\newcommand{\cU}{{\cal U}}

\newcommand{\hatbnu}{\widehat{\boldsymbol {\nu}}}
\newcommand{\vn}{ {\bf n} }
\newcommand{\uvn}{\underline{\vn}}
\newcommand{\uhatbnu}{\underline{\hatbnu}}
\newcommand{\cAb}{{\overline{\cal A}}}
\newcommand{\cFb}{{\overline{\cal F}}}
\newcommand{\cDb}{{\overline{\cal D}}}
\newcommand{\cUb}{{\overline{\cal U}}} 
\newcommand{\Boxb}{{\overline{\Box}}} 
\newcommand{\Tr}{\;{\rm Tr~}}
\newcommand{\hf}{\frac{1}{2}}
\newcommand{\qtr}{\frac{1}{4}}

\def\phib{{\overline{\phi}}}
\def\varphib{{\overline{\varphi}}}
\def\etab{{\overline{\eta}}}
\def\psib{{\overline{\psi}}}
\def\kappab{{\overline{\kappa}}}
\def\thetab{{\overline{\theta}}}
\def\chib{{\overline{\chi}}}

\def\pib{{\overline{\pi}}}
\def\lambdab{{\overline{\lambda}}}
\def\omegab{{\overline{\omega}}}
\def\sigmab{{\overline{\sigma}}}
\def\taub{{\overline{\tau}}}
\def\rhob{{\overline{\rho}}}
\def\xib{{\overline{\xi}}}

\def\nn{\nonumber}

\def\bec{\begin{center}}
\def\eec{\end{center}}
\def\beq{\begin{equation}}
\def\eeq{\end{equation}}
\def\bea{\begin{eqnarray}}
\def\eea{\end{eqnarray}}

\title{Supersymmetric quiver gauge theories on the lattice}

\preprint{DESY-13-229}

\author{Anosh Joseph \\
John von Neumann Institute for Computing NIC, Platanenallee 6, 15738 Zeuthen, Germany \\
Deutsches Elektronen-Synchrotron DESY, Platanenallee 6, 15738 Zeuthen, Germany \\ 
E-mail: \email{anosh.joseph@desy.de}}

\abstract{In this paper we detail the lattice constructions of several classes of supersymmetric quiver gauge theories in two and three Euclidean spacetime dimensions possessing exact supersymmetry at finite lattice spacing. Such constructions are obtained through the methods of topological twisting and geometric discretization of Euclidean Yang--Mills theories with eight and sixteen supercharges in two and three dimensions. We detail the lattice constructions of two-dimensional quiver gauge theories possessing four and eight supercharges and three-dimensional quiver gauge theories possessing eight supercharges.}


\begin{document}

\section{Introduction}
\label{sec:intro}

Supersymmetric quiver gauge theories participate in several interesting areas of physics, for example, in string theory and quantum integrable systems. In string theory, one can summarize the field content of the supersymmetric Yang--Mills (SYM) on the p-brane using a ``quiver diagram." One specific example of this is the role played by the three-dimensional $\cN=4$ supersymmetric quiver gauge theories in type IIB brane constructions of Hanany-Witten type \cite{Hanany:1996ie}. In the case of quantum integrable systems, one example would be the two-dimensional $\cN = (2, 2)$ supersymmetric quiver gauge theories. These theories can be related to quantum integrable systems such as spin chains through the Gauge/Bethe correspondence \cite{Nekrasov:2009ui, Nekrasov:2009uh}.

This paper is devoted to detailing the constructions of several classes of supersymmetric quiver gauge theories possessing four and eight supercharges on two- and three-dimensional Euclidean spacetime lattices. These lattice theories preserve a subset of the continuum supersymmetries exactly at finite lattice spacing. It is important to have a nonperturbative regularization of supersymmetric quiver gauge theories when we are interested in investigating the strong coupling regimes of these theories. A lattice formulation of these theories would indeed complement the search to unravel the rich structure of the above mentioned physics systems.

There has been a lot of progress in the recent past to write down the actions of supersymmetric gauge theories on a Euclidean spacetime lattice. In this work we focus on supersymmetric quiver gauge theories with extended supersymmetries on the lattice. Our starting point would be the lattice constructions of Yang--Mills theories possessing extended supersymmetries.\footnote{For a set of recent reviews see refs. \cite{Kaplan:2003uh, Giedt:2006pd, Catterall:2009it, Joseph:2011xy}.} There are two distinct formulations immediately available to us to construct extended supersymmetric Yang--Mills theories on a lattice while maintaining a subset of the continuum supersymmetries \cite{Kaplan:2002wv, Cohen:2003xe, Cohen:2003qw, Kaplan:2005ta, Catterall:2004np, Catterall:2005fd, Catterall:2006jw, Catterall:2005eh, Catterall:2013roa}. These lattice theories preserve supersymmetry exactly on the lattice, on the contrary to other approaches where supersymmetry only emerges in the continuum limit \cite{Feo:2002yi, Elliott:2005bd}. See \cite{Sugino:2003yb, Sugino:2004qd, D'Adda:2005zk, D'Adda:2007ax, Kanamori:2008bk, Hanada:2009hq, Hanada:2010kt, Hanada:2010gs, Hanada:2011qx} for other recent complementary approaches to the problem of exact lattice supersymmetry.

The first approach is known as the method of orbifolding. It is based on an orbifold projection of a supersymmetric matrix model. An appropriate projection of the matrix model generates the desired lattice theory, which preserves a subset of the supersymmetries of the target theory. The second approach, the method of topological twisting, utilizes the twists of Witten type \cite{Witten:1988ze} along with Dirac--K\"ahler fermions. The continuum action is rewritten in a twisted form and then the theory is discretized by keeping a subset of the twisted supersymmetries exact on the lattice. The twisted fermions form components of Dirac--K\"ahler fields and they have a geometric realization on the lattice as entities living on p-cells of the lattice.

We use the latter approach to construct supersymmetric quiver gauge theories on a Euclidean spacetime lattice. These two formulations appear different from the starting point but the lattices they ultimately generate are identical \cite{Unsal:2006qp, Catterall:2007kn, Catterall:2009it}. The reason for this is that in the twisting approach the fields are decomposed as representations of the twisted symmetry group, which is the diagonal subgroup of the product of the Euclidean rotation and R-symmetry groups. In the orbifold approach, the same diagonal subgroup plays a crucial role -- the orbifold projected variables are charged under this diagonal subgroup and these charges determine the placement of orbifold projected variables on the lattice. 

Though we use the method of topological twisting in writing down the lattice actions of supersymmetric quiver gauge theories in this paper, we expect that one could construct the same family of lattice theories from the method of orbifold projection.

Supersymmetric lattice gauge theories with extended supersymmetries have been constructed mostly for supersymmetric Yang--Mills theories.\footnote{It includes the well known theory, the four-dimensional $\cN=4$ SYM. It has been the subject of a few numerical studies \cite{Weir:2013zua, Catterall:2012yq, Catterall:2011aa, Mehta:2011ud, Galvez:2012sv}, with more in progress.} There have been a few extensions of these formulations by incorporating matter fields in the adjoint and fundamental representations \cite{Endres:2006ic, Giedt:2006dd, Giedt:2011zza, Matsuura:2008cfa, Joseph:2013jya}. In this paper, we detail the constructions of supersymmetric quiver lattice gauge theories in two and three dimensions possessing four and eight supercharges. These theories contain adjoint fields living on the nodes and bi-fundamental matter fields with legs placed on the nodes of the quiver. Two-dimensional quiver gauge theories with four supercharges were constructed by Matsuura in ref. \cite{Matsuura:2008cfa} while formulating four supercharge lattice gauge theories with fundamental matter. Similar construction in three dimensions has been carried out in ref. \cite{Joseph:2013jya}. There, lattice quiver gauge theories with eight supercharges have been constructed while formulating three-dimensional eight supercharge lattice gauge theories with fundamental matter fields. In this paper, we detail these quiver gauge theory constructions in a coherent way and also we complete the lattice quiver theory constructions in two and three dimensions with the addition of two-dimensional lattice quiver gauge theories possessing eight supercharges. In two dimensions, we also construct lattice quiver gauge theories with circular topology, possessing arbitrary number of nodes in the quiver.

The quiver lattice gauge theories detailed in this paper are constructed using a general procedure. We begin with a Euclidean SYM theory possessing appropriate number of supercharges. The theory is then topologically twisted to make it lattice compatible. It is dimensionally reduced to three or two dimensions in the next step. To make it a quiver gauge theory, we replicate the theory and then make an appropriate subset of the field content of the resulting theory bi-fundamental. The replicated theories form a quiver gauge theory with fields living on the nodes transforming as adjoints. The bi-fundamental fields live on the links connecting the nodes of the quiver. Changing the representation of the fields from adjoint to bi-fundamental breaks some of the supersymmetries and thus the resulting quiver gauge theories will always have lower number of supersymmetries compared to that of their parent theories.

These continuum quiver theories are then discretized on a Eulcidean spacetime lattice using the method of geometric discretization. The nodes of the quiver theory become lattice spacetimes with same dimensionalities and the bi-fundamental fields of the quiver theory have legs placed on adjacent spacetime lattices.         

The organization of this paper is as follows. In section \ref{sec:sym_twist} we provide a brief description of the method of topological twisting for SYM theories with extended supersymmetries in $d$ Euclidean spacetime dimensions. In section \ref{sec:twisted_SYM_theories} we discuss the constructions of a few classes of twisted supersymmetric Yang--Mills theories in the continuum. This include the two-dimensional $\cN = (2, 2)$ SYM, three-dimensional $\cN=4$ SYM  and four-dimensional $\cN=4$ SYM. We then detail the constructions of twisted supersymmetric quiver gauge theories in section \ref{sec:twisted_susy_quiver_theories}. The continuum twisted quiver theories constructed in this section are four and eight supercharge theories in two dimensions, including a circular quiver with eight supercharges and in three dimensions a twisted quiver theory with eight supercharges. In section \ref{sec:lattice_susy_quiver_theories} we discuss the lattice implementation supersymmetric quiver theories. We conclude with some discussions and prospects in section \ref{sec:discussions_prospects}.

\section{SYM theories and topological twisting}
\label{sec:sym_twist}

In this section we briefly discuss the twisting process of SYM theories with extended supersymmetries on flat Euclidean spacetime. The process of twisting can only be defined when the theories have Euclidean signature. We can always return to Lorentz signature, if the theory is constructed on a manifold of type $M = {\mathbb R} \times W$, by simply taking Lorentz signature on ${\mathbb R}$, which would be the cases we are focusing on in this paper.

A necessary condition for twisting is that the parent SYM theories should possess extended supersymmetries. Among the set of extended SYM theories, we focus on a special class of SYM theories that can be maximally twisted. 

SYM theories with extended supersymmetries in $d$ Euclidean spacetime dimensions contain a spacetime rotation group $SO(d)_E$ and an R-symmetry group $G_{\rm R}$. For a theory to undergo maximal twisting, its R-symmetry group must contain $SO(d)$ as a subgroup. That is,
\beq
SO(d)_E \times SO(d)_R \subset SO(d)_E \times G_{\rm R}.
\eeq
To construct the twisted theory, we embed a new rotation group $SO(d)'$ into the diagonal sum of $SO(d)_E \times SO(d)_R$ and declare this $SO(d)'$ as the new Lorentz symmetry of the theory.

After twisting, the fermions of the original theory transform as integer spin representations of the twisted rotation group $SO(d)'$. They still preserve their Grassmann odd nature but now transform as irreducible antisymmetric tensors. They can be expressed as a direct sum of p-forms with ${\rm p} = 0, \cdots, d$. The gauge bosons of the untwisted theory transform as a vector under $SO(d)'$. Among the scalars of the untwisted theory, under $SO(d)'$, $d$ of them combine to form a vector and the rest of them remain as scalars. 

The supercharges also take new forms under the twisted rotation group. They also transform like twisted fermions, in integer spin representations of the twisted rotation group. Another important feature of twisting is that in the twisted supersymmetry algebra the subalgebra containing the 0-form supercharge $\cQ$ is nilpotent
\beq
\cQ^2 = 0.
\eeq

The twisted supersymmetry algebra also implies that the momentum $P_m$, $m = 1, \cdots, d$, is the $Q$-variation of something. That is, it is $Q$-exact. This fact renders it plausible that the entire energy momentum tensor could be written in a $Q$-exact form in twisted theories. This, in turn, implies that the entire action of the theory could be expressed in a $Q$-exact form say, $S = Q \Lambda$. (In some cases, for example, the $\cN=4$ SYM in four dimensions, the twisted action can be expressed as a sum of $Q$-exact and $Q$-closed terms.) We also note that the subalgebra $Q^2 = 0$ of the twisted supersymmetry algebra does not produce any spacetime translations. This makes it possible for the twisted theory to be transported easily on to the lattice.

It should be noted that the process of twisting is just a change of variables on flat Euclidean spacetime and indeed the twisted theory is physically equivalent to its untwisted cousin.

Although the twisted formulation of supersymmetry goes back to Witten \cite{Witten:1988ze} in the construction of topological field theories, this formulation had been anticipated in earlier lattice work using Dirac--K\"ahler fields \cite{Elitzur:1982vh,Sakai:1983dg,Kostelecky:1983qu,Scott:1983ha,Aratyn:1984bc}. The precise connection between Dirac--K\"ahler fermions and topological twisting was discovered by Kawamoto and collaborators \cite{Kawamoto:1999zn,Kato:2003ss,D'Adda:2004jb}. They observed that the 0-form supercharge that arises after twisting is a nilpotent scalar and constitutes a closed subalgebra of the twisted superalgebra. It is this scalar supersymmetry that can be made manifest in the lattice theory \cite{Catterall:2004zy,Catterall:2004np,Catterall:2005fd,Catterall:2005eh,Catterall:2009it}.

In order to construct supersymmetric quiver gauge theories on the lattice we use twisted SYM theories in two, three and four Euclidean spacetime dimensions possessing four, eight and sixteen supercharges, respectively. To make our discussion more self-contained, we briefly go through the continuum twisted formulations of these theories in the next section.  

\section{Twisted SYM theories in the continuum}
\label{sec:twisted_SYM_theories}

\subsection{Two-dimensional $\cN = (2, 2)$ SYM}
\label{sec:2d_n22_twist}

The two-dimensional $\cN = (2, 2)$ Euclidean SYM can be obtained by dimensional reduction of four-dimensional $\cN = 1$ Euclidean SYM. The four-dimensional theory has global symmetry group $SO(4)_E \times U(1)$, where $SO(4)_E$ is the Euclidean Lorentz symmetry and $U(1)$ is the chiral symmetry. After dimensional reduction, the symmetry group splits into the following form
\beq
SO(2)_E \times SO(2)_{R_1} \times U(1)_{R_2}.
\eeq
Here, $SO(2)_E$ is the two-dimensional Euclidean Lorentz symmetry; $SO(2)_{R_1}$ is rotational symmetry along reduced dimensions and $U(1)_{R_2}$ is the chiral $U(1)$ symmetry of the original theory. We twist this theory by declaring a new rotational symmetry group $SO(2)'$, which is the diagonal subgroup of the product of the Lorentz rotation $SO(2)_E$ and the $SO(2)_{R_1}$ symmetry
\beq
SO(2)'={\rm diag}\Big(SO(2)_E \times SO(2)_{R_1}\Big).
\eeq

The untwisted theory contains four fermionic and four bosonic degrees of freedom. There are four real supercharges in this theory. The fermions and supersymmetries of the original theory transform as integer spin representations of the twisted rotation group. Under $SO(2)'$, the fermionic degrees of freedom of the twisted theory are p-forms with p$=0, 1, 2$. We label them as $\{\eta$, $\psi_a$, $\chi_{ab}\}$. The twisted supercharges are packaged in the set of p-forms $\{\cQ, \cQ_a, \cQ_{ab}\}$.

The two scalars of the untwisted theory combine to form a vector $B_a$ under the twisted rotation group. Since there are two vector fields in the twisted theory, $A_a$ and $B_a$, and they both transform the same way under $SO(2)'$, it is natural to combine them to form a complexified gauge field $\cA_a$ and write down the twisted theory in a compact way. Thus there are two complexified connections in the twisted theory:
\bea
\cA_a \equiv A_a + iB_a,~~~\cAb_a \equiv A_a - iB_a.
\eea

The theory contains complexified covariant derivatives and they are defined by
\bea
\label{eq:cov_der_adj1}
\cD_a ~\cdot &=& \partial_a \cdot + ~[\cA_a, ~\cdot~ ] = \partial_a \cdot + ~[A_a + iB_a, ~\cdot~ ],\\
\label{eq:cov_der_adj2}
\cDb_a ~\cdot &=& \partial_a \cdot + ~[\cAb_a, ~\cdot~ ] = \partial_a \cdot + ~[A_a - i B_a, ~\cdot~ ].
\eea
The complexification of gauge fields also results in complexified field strengths $\cF_{ab} = [\cD_a, \cD_b]$ and $\cFb_{ab} = [\cDb_a, \cDb_b]$. All fields take values in the adjoint representation of the gauge group $U(N)$. Although the theory contains complexified gauge fields and field strengths, it possesses only the usual $U(N)$ gauge-invariance corresponding to the real part of the complexified connection. 

The scalar supercharge $\cQ$ form a nilpotent subalgebra of the twisted supersymmetry algebra. It acts on the twisted fields in the following way
\begin{align}
\label{eq:susy_Q4}
\cQ \cA_a& =\psi_a,&
\cQ \cAb_a& =0,\\
\cQ \psi_a& =0,&
\cQ \chi_{ab}& =-[\cDb_a, \cDb_b],\\
\cQ \eta& =d,&
\cQ d& =0,
\end{align}
where $d$ is an auxiliary field introduced for the off-shell completion of the twisted supersymmetry algebra. It has the equation of motion
\beq
d = \sum_a [\cDb_a, \cD_a], ~a = 1, 2.
\eeq

The action of the twisted theory can be expressed in a $\cQ$-exact form
\bea
S_{{\rm SYM}}^{\cN=(2, 2), d=2} &=& \frac{1}{g_2^2} \cQ \int d^2 x \Tr \Lambda,
\eea
where $g_2$ is the coupling constant of the theory and
\bea
\Lambda &=& \chi_{ab} \cF_{ab} + \eta[\cDb_a, \cD_a] - \frac{1}{2}\eta d.
\eea

After performing the $\cQ$-variation and integrating out the field $d$ we have the action
\beq
\label{eq:2d_Q4_action}
S_{{\rm SYM}}^{\cN=(2, 2), d=2} = \frac{1}{g_2^2} \int d^2x \Tr \Big(-\cFb_{ab} \cF_{ab} + \frac{1}{2}[\cDb_a, \cD_a]^2 - \chi_{ab} \cD_a \psi_b - \eta \cDb_a \psi_a \Big).
\eeq

It is easy to see that the twisted action is $\cQ$-invariant by construction. We have $\cQ S_{{\rm SYM}}^{\cN=(2, 2), d=2} = \cQ^2 \Lambda = 0$.

\subsection{Three-dimensional $\cN=4$ SYM}
\label{sec:3d_n4_twist}

The three-dimensional Euclidean $\cN=4$ SYM can be obtained by dimensional reduction of six-dimensional Euclidean $\cN=1$ SYM. The six-dimensional theory has a gauge field and two independent Weyl spinors, with all fields transforming in the adjoint representation of the gauge group. After reducing to three dimensions the Weyl spinors split into two independent four-component complex spinors and the gauge field reduces to a three-dimensional gauge field and three real scalars. This theory contains eight real supercharges. The global symmetry group of the three-dimensional theory is $SU(2)_E \times SU(2)_R \times SU(2)_N$, where $SU(2)_E$ is the Euclidean rotation group in three dimensions, $SU(2)_R$ is the R-symmetry group of the six-dimensional theory and $SU(2)_N$ is the internal Euclidean rotation group arising from the decomposition $SO(6) \rightarrow SU(2)_E \times SU(2)_N$. 

We twist this theory by declaring a new rotation group $SU(2)^\prime$ as the diagonal subgroup of $SU(2)_E$ and $SU(2)_N$,
\beq
SU(2)'={\rm diag}\Big(SU(2)_E \times SU(2)_N\Big).
\eeq
This particular twist of the theory is known as the Blau--Thompson twist \cite{Blau:1996bx}. After twisting, the field content of the original theory becomes representations of the twisted rotation group $SU(2)'$.

The twisting process gives rise to the following spectrum of the twisted theory: a three-dimensional gauge field $A_m$, $m=1,2,3$; a vector $B_m$ composed of three scalars of the untwisted theory; and eight p-form fermions, p $= 0, 1, 2, 3$, which we conveniently represent as $\{\eta, \psi_m, \chi_{mn}, \theta_{mnr}\}$. The supercharges also undergo a decomposition similar to that of the fermions. They are packaged in the set $\{\cQ, \cQ_m, \cQ_{mn}, \cQ_{mnr}\}$.

The twisted action of the three-dimensional $\cN=4$ SYM takes the following form in the continuum
\bea
S_{{\rm SYM}}^{\cN=4, d=3} &=& S_{\cQ{\rm -exact}} + S_{\cQ{\rm -closed}},
\eea
where
\bea
S_{\cQ{\rm -exact}} &=& \frac{1}{g_3^2}\cQ \int d^3x \Tr \Big(\chi_{mn}[\cD_m, \cD_n] + \eta \left[\cDb_m,\cD_m\right] + \hf\eta d \Big),
\eea
and 
\bea
S_{\cQ{\rm -closed}} &=& - \frac{1}{g_3^2} \int d^3x \Tr \theta_{mnr}\cDb_r \chi_{mn},
\eea
with $g_3$ the coupling constant of the theory. Here also the twisted theory contains two vector fields, $A_m$ and $B_m$, and we have combined them to form a complex gauge field $\cA_m = A_m + i B_m$. Thus the degrees of freedom of the twisted theory are just the twisted fermions $\{\eta, \psi_m, \chi_{mn}, \theta_{mnr}\}$ and the complex gauge field $\cA_m$. The field $d$ is an auxiliary field introduced to render the scalar supersymmetry $\cQ$ nilpotent off-shell. It has the equation of motion 
\beq
d = \sum_m [\cDb_m, \cD_m], ~m = 1, 2, 3.
\eeq

The scalar supersymmetry acts on the twisted fields the following way
\bea
\cQ \cA_m &=& \psi_m,\\
\cQ \cAb_m &=& 0,\\
\cQ \eta &=&d,\\
\cQ d &=& 0,\\
\cQ \psi_m &=& 0,\\
\cQ \chi_{mn} &=& -[\cDb_m, \cDb_n],\\
\cQ \theta_{mnr} &=& 0.
\eea

After performing the $\cQ$-variation and integrating out the auxiliary field, the action of the three-dimensional theory becomes
\bea
\label{eq:3d_Q8_action}
S_{{\rm SYM}}^{\cN=4, d=3} &=& \frac{1}{g_3^2}\int d^3 x~\Tr \Big(-\cFb_{mn}\cF_{mn} + \hf[\cDb_m, \cD_m]^2 -\chi_{mn} \cD_m\psi_n \nn \\
&&~~~~~~~~~~~~~~~~~~~~~~~~~~~~~~~~~~~~~~~~- \psi_m\cDb_m\eta - \theta_{mnr}\cDb_r \chi_{mn}\Big).
\eea

It is easy to show that the three-dimensional twisted action is invariant under the scalar supersymmetry: $\cQ S_{{\rm SYM}}^{\cN=4, d=3} = 0$. The $\cQ$-exact piece vanishes due to the nilpotent nature of $\cQ$ and the $\cQ$-closed piece vanishes due to the Bianchi identity for the complex covariant derivatives. 

\subsection{Four-dimensional $\cN=4$ SYM}
\label{sec:4d_n4_twist}

The four-dimensional $\cN=4$ Euclidean SYM can be obtained by dimensional reduction of ten-dimensional $\cN=1$ Euclidean SYM theory. The four-dimensional theory contains four Majorana fermions, a gauge field and six scalars transforming in the adjoint representation of the gauge group. It has a Euclidean Lorentz rotation group $SO(4)_E$ and an internal symmetry group $SO(6)_R$.  We can maximally twist the four-dimensional $\cN=4$ SYM (this particular twist of the theory is known as the Marcus twist \cite{Marcus:1995mq}) to obtain a twisted theory that can be easily transported on to the lattice. The twist is carried out by declaring a new rotation group $SO(4)'$, which is the diagonal subgroup of $SO(4)_E \times SO(6)_R$ and then rewriting the fields and supersymmetries of the original theory under the twisted rotation group. 

The twisted theory contains fermions and supercharges transforming as integer spin representations of the twisted rotation group. They transform as p-forms, ${\rm p} = 0, \cdots, 4$. We can conveniently parametrize the sixteen fermions of the theory as $\{\eta, \psi_\mu, \chi_{\mu \nu}, \theta_{\mu \nu \rho}, \kappa_{\mu \nu \rho \sigma}\}$. A similar decomposition, $\{\cQ, \cQ_\mu, \cQ_{\mu \nu}, \cQ_{\mu \nu \rho}, \cQ_{\mu \nu \rho \sigma}\}$, can be applied to the sixteen supercharges of the theory.

The four gauge bosons of the untwisted theory transform as a vector, $A_{\mu}$, under the twisted rotation group. Among the six scalars, four of them are now elevated to form a vector, $B_\mu$, under $SO(4)'$. The two other scalars remain as singlets and we label them as $\phi$ and $\bar{\phi}$. 

The action of the twisted $\cN=4$ SYM can be written as a linear combination of $\cQ$-exact and $\cQ$-closed terms. We have the action
\beq
S_{{\rm SYM}}^{\cN=4, d=4} = S_{\cQ{\rm -exact}} + S_{\cQ{\rm -closed}},
\eeq
where
\bea
S_{\cQ{\rm -exact}} = \frac{1}{g_4^2}\cQ \int d^4x \Tr \Lambda,
\eea
with
\beq
\Lambda = \Big(\chi_{\mu \nu} [\cD_\mu, \cD_\nu] - \frac{1}{3!} \epsilon_{\mu \nu \lambda \rho} \theta_{\nu \lambda \rho} \cD_\mu \phi + \eta [\cDb_\mu, \cD_\mu] + \eta [\phib, \phi] - \hf \eta d\Big),
\eeq
and
\beq
S_{\cQ{\rm -closed}} = -\frac{1}{g_4^2} \int d^4x~\Tr \Big(\frac{1}{4}\epsilon_{\mu \nu \lambda \rho} \chi_{\mu \nu} [\phib, \chi_{\lambda \rho}] + \frac{1}{3!} \epsilon_{\nu \sigma \lambda \rho} \epsilon_{\nu \alpha \beta \delta} \theta_{\alpha \beta \delta} \cDb_\sigma \chi_{\lambda \rho}\Big).
\eeq
Here $g_4$ is the four-dimensional coupling and $d$ is an auxiliary field with the equation of motion
\beq
d = \sum_\mu [\cDb_\mu, \cD_\mu] + [\phib, \phi], ~\mu = 1, \cdots, 4.
\eeq

After applying the $\cQ$-variation on the $\cQ$-exact piece, the action of the theory takes the following form
\bea
\label{eq:4d_Q16_action}
S_{{\rm SYM}}^{\cN=4, d=4} &=& \frac{1}{g_4^2} \int d^4x \Tr \Big(-[\cDb_\mu, \cDb_\nu][\cD_\mu, \cD_\nu] - 2(\cDb_\mu\phib)(\cD_\mu\phi) + \hf \Big([\cDb_\mu, \cD_\mu] + [\phib, \phi]\Big)^2 \nn \\
&&~~~~- \chi_{\mu\nu} \cD_\mu \psi_\nu + 2\frac{1}{3!}\frac{1}{4!}\epsilon_{\mu \alpha \beta \delta}\epsilon_{\sigma \nu \lambda \rho} \theta_{\alpha \beta \delta} \cD_\mu \kappa_{\sigma \nu \lambda \rho} - 2\frac{1}{3!}\epsilon_{\mu\nu\lambda\rho}\theta_{\nu\lambda\rho}[\phi,\psi_\mu] \nn \\
&&~~~~- \eta \cDb_\mu \psi_\mu - \frac{1}{4!}\epsilon_{\alpha\beta\delta\sigma} \eta [\phib, \kappa_{\alpha\beta\delta\sigma}] - \qtr \epsilon_{\mu\nu\lambda\rho} \chi_{\mu\nu} [\phib, \chi_{\lambda\rho}] \nn \\
&&~~~~- \frac{1}{3!}\epsilon_{\nu\sigma\lambda\rho} \epsilon_{\nu\alpha\beta\delta} \theta_{\alpha\beta\delta} \cDb_\sigma \chi_{\lambda\rho}\Big).
\eea

The scalar supercharge acts on the twisted fields of the theory the following way
\begin{align}
\label{eq:4dsusy}
\cQ \cA_\mu & =\psi_\mu,& \cQ \psi_\mu& =0,\\
\cQ \cAb_\mu & =0,& \cQ \phib& =0,\\
\cQ \chi_{\mu \nu} & =-[\cDb_\mu, \cDb_\nu],& \cQ \phi& =\frac{1}{4!}\epsilon_{\mu \nu \lambda \rho} \kappa_{\mu \nu \lambda \rho},\\
\cQ \kappa_{\mu \nu \lambda \rho} & =0,& \cQ \theta_{\alpha \beta \delta}& =\epsilon_{\lambda \alpha \beta \delta} \cDb_\lambda \phib,\\
\cQ \eta & =d,& \cQ d& =0.
\end{align}

It can be easily shown that the twisted action is $\cQ$-invariant by construction: $\cQ S_{{\rm SYM}}^{\cN=4, d=4} = 0$. The $\cQ$-exact piece in the action vanishes due to the nilpotent nature of the scalar supercharge while the $\cQ$-closed piece of the action vanishes due to Bianchi identity.

\section{Twisted supersymmetric quiver gauge theories}
\label{sec:twisted_susy_quiver_theories}

In this section we write down the actions for twisted supersymmetric quiver gauge theories with bi-fundamental matter in two and three dimensions. Such theories can easily be obtained by dimensional reductions of the three- and four-dimensional constructions discussed in the previous section. In this section, for convenience, we label the quiver gauge theories according to their number of real supercharges $Q$.

\subsection{Two-dimensional twisted quiver gauge theories}
\label{sec:2d_quiver_theories}

\subsubsection{The $Q = 4$ quiver gauge theory}
\label{sec:2d_Q4_quiver}

We can dimensionally reduce the eight supercharge theory in three dimensions, given in section (\ref{sec:3d_n4_twist}), to obtain an eight supercharge theory with adjoint matter in two dimensions. We obtain the following form of the action for the two-dimensional theory after dimensionally reducing eq. (\ref{eq:3d_Q8_action}) to two dimensions,
\beq
\label{eq:2d_Q8_adj}
S = S_{{\rm SYM}}^{\cN=(2, 2), d=2} + S_{{\rm adj~matter}},
\eeq
where the first piece of the action is the twisted action of the two-dimensional $\cN=(2, 2)$ SYM given in eq. (\ref{eq:2d_Q4_action}). The second piece includes matter fields $\{\phi$, $\phib$, $\etab$, $\psib_a$, $\chib_{ab}\}$ in the adjoint representation of the gauge group $U(N)$. It is given by
\bea
S_{{\rm adj~matter}} &=& \frac{1}{g_2^2} \int d^2 x~\Tr \Big(-2(\cDb_a\phib)(\cD_a\phi) + [\cDb_a, \cD_a][\phib, \phi] - \chib_{ab}\cDb_b \psib_a \nn \\
&&~~~~+ \etab \cD_a \psib_a - \psi_a [\phi, \psib_a] - \eta [\phib, \etab] - \hf \chi_{ab} [\phib, \chib_{ab}] + \hf [\phib, \phi]^2 \Big).
\eea

The fields of this two-dimensional theory respect the following scalar supersymmetry transformations
\begin{align}
\label{eq:susy_adj_2d_Q8}
\cQ \cA_a& =\psi_a,&
\cQ \cAb_a& =0,\\
\cQ \psi_a& =0,&
\cQ \chi_{ab}& =-[\cDb_a, \cDb_b],\\
\cQ \eta& = [\cDb_a, \cD_a] + [\phib, \phi],&
\cQ \chib_{ab} &=0,\\
\cQ \phi& =\etab,&
\cQ \phib& =0,\\
\cQ \etab& =0,&
\cQ \psib_a& = \cDb_a \phib.
\end{align}

\begin{figure}
\begin{center}
\includegraphics[width=0.3\textwidth]{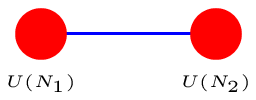}
\end{center}
\caption{\label{fig:doubleq}A quiver diagram for $U(N_1) \times U(N_2)$ gauge theory. The $U(N_1)$ and $U(N_2)$ theories interact via bi-fundamental fields (the blue line). The adjoint fields of the theory live on the nodes of the quiver (the red disks).}
\end{figure}

We can construct a two-dimensional quiver gauge theory with $Q=4$ supersymmetry from the above theory. We are interested in constructing a quiver gauge theory with two nodes. For that we double the number of fields of the original theory and then change the representations of a subset of the fields from adjoint to bi-fundamental. This results in two interacting gauge theories with a product gauge group $U(N_1) \times U(N_2)$ possessing $Q=4$ supersymmetry. The SYM multiplets of this quiver gauge theory transform in the adjoint representation of the product gauge group and they live on the nodes of the quiver. The two theories interact via matter multiplets in the bi-fundamental representation of $U(N_1) \times U(N_2)$. (See figure \ref{fig:doubleq}.) The action of the four supercharge quiver gauge theory can be decomposed in the following way
\beq
S_{{\rm quiver}}^{Q=4, d=2} = S^{\rm SYM}_{({\bf adj}, {\bf 1})} + S^{\rm SYM}_{({\bf 1}, {\bf adj})} + S^{\rm matter}_{(\Box, \Boxb)} + S^{\rm matter}_{(\Boxb, \Box)},
\eeq
with the field content of the quiver theory $\{\cA_a$, $\cAb_a$, $\eta$, $\psi_a$, $\chi_{ab}\}$, $\{\widehat{\cA}_a$, $\widehat{\cAb}_a$, $\widehat{\eta}$, $\widehat{\psi}_a$, $\widehat{\chi}_{ab}\}$, $\{\phi$, $\widehat{\phib}$, $\etab$, $\widehat{\psib}_a$, $\chib_{ab}\}$ and $\{\widehat{\phi}$, $\phib$, $\widehat{\etab}$, $\psib_a$, $\widehat{\chib}_{ab}\}$ transforming respectively as $({\bf adj}, {\bf 1})$, $({\bf 1}, {\bf adj})$, $(\Box, \Boxb)$ and $(\Boxb, \Box)$ under the gauge group $U(N_1) \times U(N_2)$. 

The pieces of the four supercharge quiver gauge theory action are given below
\beq
S^{\rm SYM}_{({\bf adj}, {\bf 1})} = \frac{1}{g_2^2} \int d^2x \Tr \Big(-\cFb_{ab} \cF_{ab} + \frac{1}{2}[\cDb_a, \cD_a]^2 - \chi_{ab} \cD_a\psi_b - \eta \cDb_a \psi_a \Big),
\eeq

\beq
S^{\rm SYM}_{({\bf 1}, {\bf adj})} = \frac{1}{g_2^2} \int d^2x \Tr \Big(-\widehat{\cFb}_{ab} \widehat{\cF}_{ab} + \frac{1}{2}[\widehat{\cDb}_a, \widehat{\cD}_a]^2 - \widehat{\chi}_{ab} \widehat{\cD}_a \widehat{\psi}_b - \widehat{\eta} \widehat{\cDb}_a \widehat{\psi}_a \Big),
\eeq

\bea
S^{\rm matter}_{(\Box, \Boxb)} &=& \frac{1}{g_2^2} \int d^2 x~\Tr \Big(2 \widehat{\phib} \cDb_a \cD_a \widehat{\phi} + [\cDb_a, \cD_a] (\widehat{\phib} \widehat{\phi} - \phi \phib) - \chib_{ab} \cDb_b \psib_a \nn \\
&&~~~~~~~~~~+ \widehat{\psib}_a \cD_a \widehat{\etab} - \psi_a (\phi \psib_a - \widehat{\psib}_a \widehat{\phi}) - \eta (\widehat{\phib} \widehat{\etab} - \etab \phib) \nn \\
&&~~~~~~~~~~- \hf \chi_{ab} (\widehat{\phib} \widehat{\chib}_{ab} \nn - \chib_{ab} \phib) + \hf (\widehat{\phib} \widehat{\phi} - \phi \phib)^2 \Big)
\eea

and

\bea
S^{\rm matter}_{(\Boxb, \Box)} &=&\frac{1}{g_2^2} \int d^2 x~\Tr \Big(2 \phib \cDb_a \cD_a \phi + [\cDb_a, \cD_a] (\phib \phi - \widehat{\phi} \widehat{\phib}) - \widehat{\chib}_{ab} \cDb_b \widehat{\psib}_a \nn \\
&&~~~~~~~~~~+ \psib_a \cD_a \etab - \widehat{\psi}_a (\widehat{\phi} \widehat{\psib}_a - \psib_a \phi) - \widehat{\eta} (\phib \etab - \widehat{\etab} \widehat{\phib}) \nn \\
&&~~~~~~~~~~- \hf \widehat{\chi}_{ab} (\phib \chib_{ab} - \widehat{\chib}_{ab} \widehat{\phib}) + \hf (\phib \phi - \widehat{\phi} \widehat{\phib})^2 \Big).
\eea

There are two types of covariant derivatives appearing in the above expressions, acting respectively on adjoint and bi-fundamental fields. The covariant derivatives for the adjoint fields are given in eqs. (\ref{eq:cov_der_adj1}) - (\ref{eq:cov_der_adj2}). For a generic bi-fundamental field $\phi$ in the representation $(\Box, \Boxb)$ we have the action of the covariant derivative
\beq
\cD_a \phi = \partial_a \phi + \cA_a \phi - \phi \widehat{\cA}_a,
\eeq
with $\cA_a$ and $\widehat{\cA}_a$ the gauge fields for $U(N_1)$ and $U(N_2)$ respectively. The gauge transformation rule for the field $\phi$, under $(G, H) \in U(N_1) \times U(N_2)$, is given by $\phi \rightarrow G \phi H^\dagger$. For a field $\widehat{\phi}$ in the representation $(\Boxb, \Box)$ we have the action of the covariant derivative
\beq
\cD_a \widehat{\phi} = \partial_a \widehat{\phi} + \widehat{\cA}_a \widehat{\phi} - \widehat{\phi} \cA_a,
\eeq
with the rule for gauge transformation: $\widehat{\phi} \rightarrow H \widehat{\phi} G^\dagger$.

\subsubsection{The $Q = 8$ quiver gauge theory}
\label{sec:2d_Q8_quiver}

To construct a quiver gauge theory with $Q=8$ supersymmetry in two dimensions we begin by dimensionally reducing the four-dimensional sixteen supercharge twisted SYM given in section (\ref{sec:4d_n4_twist}). We obtain the following form of the action for the two-dimensional sixteen supercharge theory after dimensional reduction 
\beq
\label{eq:2d_Q16_adj}
S = S_{{\rm SYM}}^{\cN=(4, 4), d=2} + S_{{\rm adj~matter}},
\eeq
where the first piece of the action is the twisted action of the two-dimensional $\cN=(4, 4)$ SYM. It is
\bea
S_{{\rm SYM}}^{\cN=(4, 4), d=2} &=& \frac{1}{g_2^2} \int d^2 x~\Tr \Big(- \cFb_{ab} \cF_{ab} + \hf [\cDb_a, \cD_a]^2 - \eta \cDb_a \psi_a - \chi_{ab} \cD_a \psi_b \nn \\
&&~~~~- 2 (\cDb_a \pib)(\cD_a \pi) + [\cDb_a, \cD_a][\pib, \pi] + \lambdab \cD_a \omegab_a - \sigmab_{ab}\cDb_b \omegab_a \nn \\
&&~~~~- \eta [\pib, \lambdab] - \psi_a [\pi, \omegab_a] - \hf \chi_{ab} [\pib, \sigmab_{ab}] + \hf [\pib, \pi]^2 \Big).
\eea
It contains two complexified bosons $\{\cA_a, \cAb_a\}$, two scalars $\{\pi, \pib\}$ and eight twisted fermions $\{\eta, \psi_a, \chi_{ab}, \lambdab, \omegab_a, \sigmab_{ab}\}$ transforming in the adjoint representation of gauge group $U(N)$. Note that this piece of the action can be obtained from dimensionally reducing the three-dimensional eight supercharge theory, given in section (\ref{sec:3d_n4_twist}), down to two dimensions.

The second piece includes matter fields $\{\phi$, $\phib$, $\varphi$, $\varphib$, $\etab$, $\psib_a$, $\kappab_{ab}$, $\rhob$, $\taub_a$, $\xib_{ab}\}$ in the adjoint representation of the gauge group $U(N)$. The matter action is
\bea
\label{eq:3d_q16_matter}
 S_{{\rm adj~matter}} &=& \frac{1}{g_2^2} \int d^2 x~\Tr \Big[-2 (\cDb_a\varphib)(\cD_a\varphi) - 2(\cDb_a\phib)(\cD_a\phi) \nn \\
&&~~~~+ [\cDb_a, \cD_a]([\varphib, \varphi]+ [\phib, \phi]) \nn \\
&&~~~~+ \etab \cD_a \psib_a - \rhob \cDb_a \taub_a - \kappab_{ab} \cDb_a \psib_b - 2 \xib_{ab} \cD_a \taub_b \nn \\
&&~~~~- \eta [\phib, \etab] - 3\epsilon_{ab} \eta [\varphib, \xib_{ab}] + \rhob [\pi, \etab] + \taub_a [\pib, \psib_a] + \hf \xib_{ab}[\pi, \kappab_{ab}] \nn \\
&&~~~~+ \frac{3}{2} \epsilon_{ab} \sigmab_{ab} [\varphi, \etab] - \psi_a [\phi, \psib_a] - \lambdab [\phi, \rhob] - \hf \epsilon_{ab} \lambdab [\varphi, \kappab_{ab}] \nn \\
&&~~~~+ \epsilon_{ab} \psi_a[\varphi, \taub_b] + 2\epsilon_{ab} \omegab_a [\varphib, \psib_b] + \epsilon_{ab} \chi_{ab}[\varphib, \rhob] \nn \\
&&~~~~- \hf \chi_{ab} [\phib, \kappab_{ab}] - \omegab_a [\phib, \taub_a] - \frac{3}{2} \sigmab_{ab} [\phi, \xib_{ab}] \nn \\
&&~~~~-2 [\pib, \varphib][\pi, \varphi]  - 2[\pib, \phib][\pi, \phi] - 2[\varphib, \phib][\varphi, \phi] \nn \\
&&~~~~+ \hf \Big([\varphib, \varphi]+ [\phib, \phi]\Big)^2 + [\pib, \pi] ([\varphib, \varphi] + [\phib, \phi]) \Big].
\eea

The fields of this two-dimensional sixteen supercharge theory respect the following scalar supersymmetry transformations
\begin{align}
\label{eq:susy_adj_2d_Q16}
\cQ \cA_a& =\psi_a,&
\cQ \cAb_a& =0,\\
\cQ \eta& =[\cDb_a, \cD_a] + [\pib, \pi] + [\varphib, \varphi] + [\phib, \phi],&
\cQ \psi_a& =0,\\
\cQ \chi_{ab} & =-[\cDb_a, \cDb_b],&
\cQ \varphi& = \hf \epsilon_{ab}\xib_{ab},\\
\cQ \varphib& =0,&
\cQ \phi& =\etab,\\
\cQ \phib& =0,&
\cQ \pi& =\lambdab,\\
\cQ \pib& =0,&
\cQ \lambdab& =0,\\
\cQ \etab& =0,&
\cQ \rhob& =[\pib, \phib],\\
\cQ \omegab_a& =\cDb_a\pib,&
\cQ \psib_a& =\cDb_a\phib,\\
\cQ \taub_a & = \epsilon_{ab}\cDb_b\varphib,&
\cQ \sigmab_{ab}& = 2\epsilon_{ab} [\varphib, \phib],\\
\cQ \kappab_{ab}& = \epsilon_{ab} [\pib, \varphib],&
\cQ \xib_{ab}& =0.
\end{align}

In order to construct a two-dimensional quiver gauge theory with $Q=8$ supersymmetry we replicate the action given in eq. (\ref{eq:2d_Q16_adj}) and make a subset of the fields bi-fundamental. In this case, we have sufficient number of fields and supersymmetries to construct different types of quiver theories. We begin with the construction of a quiver theory with two nodes. The procedure is the same as that of the two-dimensional four supercharge quiver theory construction detailed in the previous section. Thus we have two interacting gauge theories with gauge group $U(N_1) \times U(N_2)$.  The quiver theory has adjoint fields living on the nodes and bi-fundamental fields living on the links connecting the nodes. The action of the quiver gauge theory is
\beq
S_{{\rm quiver}}^{Q=8, d=2} = S^{\rm SYM}_{({\bf adj}, {\bf 1})} + S^{\rm SYM}_{({\bf 1}, {\bf adj})} + S^{\rm matter}_{(\Box, \Boxb)} + S^{\rm matter}_{(\Boxb, \Box)}.
\eeq
The quiver theory contains the following sets of fields grouped according to their transformation properties under $U(N_1) \times U(N_2)$. The set of fields $\{\cA_m$, $\cAb_m$, $\pi$, $\pib$, $\eta$, $\psi_m$, $\chi_{mn}$, $\lambdab$, $\omegab_a$, $\sigmab_{ab}\}$ transforms as $({\bf adj}, {\bf 1})$, $\{\widehat{\cA}_m$, $\widehat{\cAb}_m$, $\widehat{\pi}$, $\widehat{\pib}$, $\widehat{\eta}$, $\widehat{\psi}_m$, $\widehat{\chi}_{mn}$, $\widehat{\lambdab}$, $\widehat{\omegab}_a$, $\widehat{\sigmab}_{ab}\}$ transforms as $({\bf 1}, {\bf adj})$, $\{\widehat{\phi}$, $\phib$, $\varphi$, $\widehat{\varphib}$, $\widehat{\etab}$, $\psib_a$, $\widehat{\kappab}_{ab}$, $\rhob$, $\widehat{\taub}_a$, $\xib_{ab}\}$ transforms as $(\Box, \Boxb)$ and $\{\phi$, $\widehat{\phib}$, $\widehat{\varphi}$, $\varphib$, $\etab$, $\widehat{\psib}_a$, $\kappab_{ab}$, $\widehat{\rhob}$, $\taub_a$, $\widehat{\xib}_{ab}\}$ transforms as $(\Boxb, \Box)$ under the gauge group $U(N_1) \times U(N_2)$.

The pieces of the quiver gauge theory action are given below
\bea
S^{\rm SYM}_{({\bf adj}, {\bf 1})} &=& \frac{1}{g_2^2} \int d^2 x~\Tr \Big(- \cFb_{ab} \cF_{ab} + \hf [\cDb_a, \cD_a]^2 - \eta \cDb_a \psi_a - \chi_{ab} \cD_a \psi_b \nn \\
&&~~~~- 2 (\cDb_a \pib)(\cD_a \pi) + [\cDb_a, \cD_a][\pib, \pi] + \lambdab \cD_a \omegab_a - \sigmab_{ab}\cDb_b \omegab_a \nn \\
&&~~~~- \eta [\pib, \lambdab] - \psi_a [\pi, \omegab_a] - \hf \chi_{ab} [\pib,\sigmab_{ab}] + \hf [\pib, \pi]^2 \Big),
\eea

\bea
S^{\rm SYM}_{({\bf 1}, {\bf adj})} &=& \frac{1}{g_2^2} \int d^2 x~\Tr \Big(- \widehat{\cFb}_{ab} \widehat{\cF}_{ab} + \hf [\widehat{\cDb}_a, \widehat{\cD}_a]^2 - \widehat{\eta} \widehat{\cDb}_a \widehat{\psi}_a - \widehat{\chi}_{ab} \widehat{\cD}_a \widehat{\psi}_b \nn \\
&&~~~~- 2 (\widehat{\cDb}_a \widehat{\pib})(\widehat{\cD}_a \widehat{\pi}) + [\widehat{\cDb}_a, \widehat{\cD}_a][\widehat{\pib}, \widehat{\pi}] + \widehat{\lambdab} \widehat{\cD}_a \widehat{\omegab}_a - \widehat{\sigmab}_{ab} \widehat{\cDb}_b \widehat{\omegab}_a \nn \\
&&~~~~- \widehat{\eta} [\widehat{\pib}, \widehat{\lambdab}] - \widehat{\psi}_a [\widehat{\pi}, \widehat{\omegab}_a] - \hf \widehat{\chi}_{ab} [\widehat{\pib}, \widehat{\sigmab}_{ab}] + \hf [\widehat{\pib}, \widehat{\pi}]^2 \Big),
\eea

\bea
S^{\rm matter}_{(\Box, \Boxb)} &=& \frac{1}{g_2^2} \int d^2 x~\Tr \Big[2 \widehat{\varphib} \cDb_a\cD_a \widehat{\varphi} + 2 \phib \cDb_a\cD_a \phi \nn \\
&&~~~~+ [\cDb_a, \cD_a] (\widehat{\varphib} \widehat{\varphi} - \varphi \varphib + \phib \phi - \widehat{\phi} \widehat{\phib}) \nn \\
&&~~~~+ \etab \cD_a \psib_a - \widehat{\rhob} \cDb_a \widehat{\taub}_a - \widehat{\kappab}_{ab} \cDb_a \widehat{\psib}_b - 2 \xib_{ab} \cD_a \taub_b \nn \\
&&~~~~+ \rhob (\widehat{\pi} \etab - \etab \pi) + \widehat{\taub}_a (\widehat{\pib} \widehat{\psib}_a - \widehat{\psib}_a \pib) + \hf \xib_{ab} (\widehat{\pi} \kappab_{ab} - \kappab_{ab} \pi) \nn \\
&&~~~~- \eta (\phib \etab - \widehat{\etab} \widehat{\phib}) - 3\epsilon_{ab} \eta (\widehat{\varphib} \widehat{\xib}_{ab} - \xib_{ab} \varphib) + \frac{3}{2} \epsilon_{ab} \sigmab_{ab} (\varphi \etab - \widehat{\etab} \widehat{\varphi}) \nn \\
&&~~~~- \psi_a (\widehat{\phi} \widehat{\psib}_a - \psib_a \phi) - \lambdab (\widehat{\phi} \widehat{\rhob} - \rhob \phi) - \hf \epsilon_{ab} \lambdab (\varphi \kappab_{ab} - \widehat{\kappab}_{ab} \widehat{\varphi}) \nn \\
&&~~~~- \epsilon_{ab} \psi_b (\varphi \taub_a - \widehat{\taub}_a \widehat{\varphi}) - 2\epsilon_{ab} \omegab_b (\widehat{\varphib} \widehat{\psib}_a - \psib_a \varphib) \nn \\
&&~~~~+ \epsilon_{ab} \chi_{ab} (\widehat{\varphib} \widehat{\rhob} - \rhob \varphib) - \hf \chi_{ab}(\phib \kappab_{ab} - \widehat{\kappab}_{ab}\widehat{\phib}) - \omegab_a(\phib \taub_a - \widehat{\taub}_a \widehat{\phib}) \nn \\
&&~~~~- \frac{3}{2} \sigmab_{ab}(\widehat{\phi} \widehat{\xib}_{ab} - \xib_{ab}\phi) -2 (\pi \varphi - \varphi \widehat{\pi}) (\widehat{\pib} \varphib - \varphib \pib) \nn \\
&&~~~~- 2(\pib \phib - \phib \widehat{\pib} )(\widehat{\pi} \phi - \phi \pi ) - 2(\widehat{\varphib} \widehat{\phib} - \phib \varphib )(\varphi \phi - \widehat{\phi} \widehat{\varphi} ) \nn \\
&&~~~~+ \hf \Big(\widehat{\varphib} \widehat{\varphi} - \varphi \varphib + \phib \phi - \widehat{\phi} \widehat{\phib}\Big)^2 + [\pib, \pi] \Big(\widehat{\varphib} \widehat{\varphi} - \varphi \varphib + \phib \phi - \widehat{\phi} \widehat{\phib}\Big) \Big],
\eea

and

\bea
S^{\rm matter}_{(\Boxb, \Box)} &=& \frac{1}{g_2^2} \int d^2 x~\Tr \Big[2 \varphib \cDb_a\cD_a \varphi + 2 \widehat{\phib} \cDb_a\cD_a \widehat{\phi} \nn \\
&&~~~~+ [\cDb_a, \cD_a] (\varphib \varphi - \widehat{\varphi} \widehat{\varphib} + \widehat{\phib} \widehat{\phi} - \phi \phib) \nn \\
&&~~~~+ \widehat{\etab} \cD_a \widehat{\psib}_a - \rhob \cDb_a \taub_a - \kappab_{ab} \cDb_a \psib_b - 2 \widehat{\xib}_{ab} \cD_a \widehat{\taub}_b \nn \\
&&~~~~+ \widehat{\rhob} (\pi \widehat{\etab} - \widehat{\etab} \widehat{\pi}) + \taub_a (\pib \psib_a - \psib_a \widehat{\pib}) + \hf \widehat{\xib}_{ab} (\pi \widehat{\kappab}_{ab} - \widehat{\kappab}_{ab} \widehat{\pi}) \nn \\
&&~~~~- \widehat{\eta} (\widehat{\phib} \widehat{\etab} - \etab \phib) - 3\epsilon_{ab} \widehat{\eta} (\varphib \xib_{ab} - \widehat{\xib}_{ab} \widehat{\varphib}) + \frac{3}{2} \epsilon_{ab} \widehat{\sigmab}_{ab} (\widehat{\varphi} \widehat{\etab} - \etab \varphi) \nn \\
&&~~~~- \widehat{\psi}_a (\phi \psib_a - \widehat{\psib}_a \widehat{\phi}) - \widehat{\lambdab} (\phi \rhob - \widehat{\rhob} \widehat{\phi}) - \hf \epsilon_{ab} \widehat{\lambdab} (\widehat{\varphi} \widehat{\kappab}_{ab} - \kappab_{ab} \varphi) \nn \\
&&~~~~- \epsilon_{ab} \widehat{\psi}_b (\widehat{\varphi} \widehat{\taub}_a - \taub_a \varphi) - 2\epsilon_{ab} \widehat{\omegab}_b (\varphib \psib_a - \widehat{\psib}_a \widehat{\varphib}) \nn \\
&&~~~~+ \epsilon_{ab} \widehat{\chi}_{ab} (\varphib \rhob - \widehat{\rhob} \widehat{\varphib}) - \hf \widehat{\chi}_{ab}(\widehat{\phib} \widehat{\kappab}_{ab} - \kappab_{ab} \phib) - \widehat{\omegab}_a(\widehat{\phib} \widehat{\taub}_a - \taub_a \phib) \nn \\
&&~~~~- \frac{3}{2} \widehat{\sigmab}_{ab}(\phi \xib_{ab} - \widehat{\xib}_{ab} \widehat{\phi}) -2 (\widehat{\pi} \widehat{\varphi} - \widehat{\varphi} \pi) (\pib \widehat{\varphib} - \widehat{\varphib} \widehat{\pib}) \nn \\
&&~~~~- 2(\widehat{\pib} \widehat{\phib} - \widehat{\phib} \pib)(\pi \widehat{\phi} - \widehat{\phi} \widehat{\pi}) - 2(\varphib \phib - \widehat{\phib} \widehat{\varphib})(\widehat{\varphi} \widehat{\phi} - \phi \varphi) \nn \\
&&~~~~+ \hf \Big(\varphib \varphi - \widehat{\varphi} \widehat{\varphib} + \widehat{\phib} \widehat{\phi} - \phi \phib \Big)^2 + [\widehat{\pib}, \widehat{\pi}] \Big(\varphib \varphi - \widehat{\varphi} \widehat{\varphib} + \widehat{\phib} \widehat{\phi} - \phi \phib \Big) \Big].
\eea

\subsubsection{The $Q = 8$ circular quiver gauge theory}
\label{sec:2d_Q8_circular_q}

\begin{figure}
\begin{center}
\includegraphics[width=0.4\textwidth]{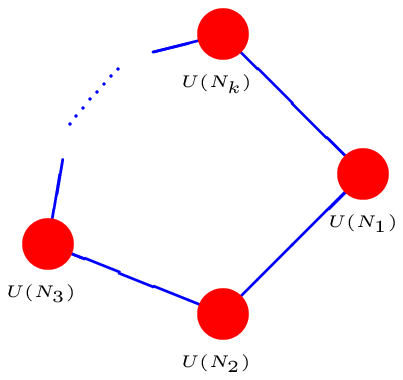}
\end{center}
\caption{\label{fig:circularq}A quiver gauge theory with circular topology. The theory contains $k$ nodes with gauge group $U(N_i)$, $1 \leq i \leq k$ with the identification $U(N_{k+1}) \equiv U(N_1)$. The $U(N_i)$ theories interact via bi-fundamental fields (the blue lines) which form a circular chain. The adjoint fields of the theory live on the nodes of the quiver (the red disks).}
\end{figure}

We can extend the two-dimensional $Q=8$ quiver gauge theory discussed in the previous section to a quiver theory with circular topology. In this case, the gauge group of the theory is $U(N_1) \times U(N_2) \times \cdots \times U(N_k)$ with $U(N_{k+1})$ identified with $U(N_1)$. The SYM multiplets of this quiver theory transform in the adjoint representation of the gauge group $U(N_1) \times U(N_2) \times \cdots \times U(N_k)$. The matter fields of this theory transform in the bi-fundamental representation of each consecutive pair of the gauge groups $U(N_i) \times U(N_{i+1})$ where $1 \leq i \leq k$ with the convention that $U(N_{k+1}) \equiv U(N_1)$. (See figure \ref{fig:circularq}.)

The construction of a circular quiver gauge theory with $Q=8$ supersymmetry is as follows. On the node $i$ of the quiver theory we place the adjoint fields $\{\cA_m$, $\cAb_m$, $\pi$, $\pib$, $\eta$, $\psi_m$, $\chi_{mn}$, $\lambdab$, $\omegab_a$, $\sigmab_{ab}\}$. On the node $i+1$ we place the adjoint fields $\{\widehat{\cA}_m$, $\widehat{\cAb}_m$, $\widehat{\pi}$, $\widehat{\pib}$, $\widehat{\eta}$, $\widehat{\psi}_m$, $\widehat{\chi}_{mn}$, $\widehat{\lambdab}$, $\widehat{\omegab}_a$, $\widehat{\sigmab}_{ab}\}$. On the node $i-1$ we place the following set of adjoint fields $\{\dot{\cA}_m$, $\dot{\cAb}_m$, $\dot{\pi}$, $\dot{\pib}$, $\dot{\eta}$, $\dot{\psi}_m$, $\dot{\chi}_{mn}$, $\dot{\lambdab}$, $\dot{\omegab}_a$, $\dot{\sigmab}_{ab}\}$. The bi-fundamental fields connecting the adjacent nodes are distributed in the following way. For the nodes $i$ and $i+1$ we have the bi-fundamental fields $\{\phi$, $\widehat{\phib}$, $\etab$, $\widehat{\psib}_a$, $\kappab_{ab}\}$ transforming in the representation $(\Boxb, \Box)$ while the set of fields $\{\widehat{\phi}$, $\phib$, $\widehat{\etab}$, $\psib_a$, $\widehat{\kappab}_{ab}\}$ transforming in the representation $(\Box, \Boxb)$. The bi-fundamental fields connecting the nodes $i$ and $i-1$ are distributed in the following way. The set of bi-fundamentals $\{\varphi$, $\dot{\varphib}$, $\rhob$, $\dot{\taub}_a$, $\xib_{ab}\}$ transforms as $(\Boxb, \Box)$ while the set of fields $\{\dot{\varphi}$, $\varphib$, $\dot{\rhob}$, $\taub_a$, $\dot{\xib}_{ab}\}$ transforms as $(\Box, \Boxb)$.

\subsection{Three-dimensional $Q=8$ quiver gauge theory}
\label{sec:3d_q8_quiver_theory}

To construct a quiver gauge theory with $Q=8$ supersymmetry in three dimensions we begin by dimensionally reducing the four-dimensional sixteen supercharge twisted SYM given in section (\ref{sec:4d_n4_twist}). We obtain the following form of the action for the three-dimensional theory after dimensional reduction
\beq
\label{eq:3d_Q8_adj}
S = S_{{\rm SYM}}^{\cN=4, d=3} + S_{{\rm adj~matter}},
\eeq
where the first piece of the action is the twisted action of the three-dimensional $\cN=4$ SYM. It is
\bea
\label{eq:3d_q8_SYM}
S_{\rm SYM}^{\cN=4, d=3} &=& \frac{1}{g_3^2} \int d^3x \Tr \Big(-\cFb_{mn}\cF_{mn} + \hf [\cDb_m, \cD_m]^2 - \eta \cDb_m \psi_m \nn \\
&&~~~~- \chi_{mn} \cD_m \psi_n - \theta_{mnr} \cDb_r \chi_{mn} \Big).
\eea

The adjoint matter part of the action is given by
\bea
S_{{\rm adj~matter}}  &=& \frac{1}{g_3^2} \int d^3x \Tr \Big[-2 (\cDb_m\varphib)(\cD_m\varphi) - 2(\cDb_m\phib)(\cD_m\phi) \nn \\
&&~~~~+ [\cDb_m, \cD_m]\Big([\varphib, \varphi]+ [\phib, \phi]\Big) \nn \\
&&~~~~+ \etab \cD_m \psib_m + \kappab_{mn} \cDb_m \psib_n + \hf \thetab_{mnr} \cD_m \kappab_{nr} \nn \\
&&~~~~- \hf \epsilon_{mnr} \psi_r [\varphi, \kappab_{mn}] + \epsilon_{mnr}\theta_{mnr} [\varphi, \etab] - \psi_m [\phi, \psib_m] - \theta_{mnr} [\phi, \thetab_{mnr}] \nn \\
&&~~~~- \epsilon_{mnr} \eta [\varphib, \thetab_{mnr}] - \eta [\phib, \etab] - \hf \chi_{mn} [\phib, \kappab_{mn}] + \epsilon_{mnr} \chi_{mn}[\varphib, \psib_r]  \nn \\
&&~~~~+ \hf \Big([\varphib, \varphi]+ [\phib, \phi]\Big)^2 - 2[\varphib, \phib][\varphi, \phi]\Big].
\eea

The fields of this three-dimensional theory respect the following scalar supersymmetry transformations

\begin{align}
\label{eq:susy_adj_3d_Q8}
\cQ \cA_m& =\psi_m,&
\cQ \cAb_m& =0,\\
\cQ \eta& =[\cDb_m, \cD_m] + [\varphib, \varphi] + [\phib, \phi],&
\cQ \psi_m& =0,\\
\cQ \chi_{mn} & =-[\cDb_m, \cDb_n],&
\cQ \theta_{mnr} & = \epsilon_{mnr}[\varphib, \phib],\\
\cQ \phi& =\etab,&
\cQ \phib& =0,\\
\cQ \varphi& =\frac{1}{3!}\epsilon_{mnr}\thetab_{mnr},&
\cQ \varphib& =0,\\
\cQ \etab& =0,&
\cQ \psib_m& = \cDb_m\phib,\\
\label{eq:susy-3d-2}
\cQ \kappab_{mn} & = \epsilon_{mnr} \cDb_r\varphib,&
\cQ \thetab_{mnr}& =0.
\end{align}

We can rewrite the above action, such that the theory becomes a three-dimensional quiver gauge theory with $\cN=4$ supersymmetry. There are two interacting $U(N)$ gauge theories within the quiver. The SYM multiplets of this quiver gauge theory transform in the adjoint representation of the gauge group $U(N_1) \times U(N_2)$. The two theories interact via matter multiplets in the bi-fundamental representation of $U(N_1) \times U(N_2)$. The action of the quiver gauge theory has the following form
\beq
S = S^{\rm SYM}_{({\bf adj}, {\bf 1})} + S^{\rm SYM}_{({\bf 1}, {\bf adj})} + S^{\rm matter}_{(\Box, \Boxb)} + S^{\rm matter}_{(\Boxb, \Box)},
\eeq
where
\bea
S^{\rm SYM}_{({\bf adj}, {\bf 1})} &=& \frac{1}{g_3^2} \int d^3x \Tr \Big(-\cFb_{mn} \cF_{mn} + \hf [\cDb_m, \cD_m]^2 - \eta \cDb_m \psi_m \nn \\
&&~~~~~~~~~~~~~~~~~~- \chi_{mn} \cD_m\psi_n - \theta_{mnr}\cDb_r \chi_{mn} \Big),
\eea

\bea
S^{\rm SYM}_{({\bf 1}, {\bf adj})} &=& \frac{1}{g_3^2} \int d^3x \Tr \Big(-\widehat{\cFb}_{mn} \widehat{\cF}_{mn} + \hf [\widehat{\cDb}_m, \widehat{\cD}_m]^2 - \widehat{\eta} \widehat{\cDb}_m \widehat{\psi}_m \nn \\
&&~~~~~~~~~~~~~~~~~~- \widehat{\chi}_{mn} \widehat{\cD}_m\widehat{\psi}_n - \widehat{\theta}_{mnr}\widehat{\cDb}_r \widehat{\chi}_{mn} \Big),
\eea

\bea
S^{\rm matter}_{(\Box, \Boxb)}  &=& \frac{1}{g_3^2} \int d^3x \Tr \Big[2 \widehat{\varphib} \cDb_m \cD_m \widehat{\varphi} + 2 \phib \cDb_m \cD_m \phi + [\cDb_m, \cD_m] \Big(\widehat{\varphib} \widehat{\varphi} - \varphi \varphib + \phib \phi  - \widehat{\phi} \widehat{\phib} \Big) \nn \\
&&~~~~+ \etab \cD_m \psib_m - \widehat{\kappab}_{np} \cDb_p \widehat{\psib}_n + \hf \thetab_{mnr} \cD_m \kappab_{nr} \nn \\
&&~~~~- \hf \epsilon_{mnr}\psi_r (\varphi \kappab_{mn} - \widehat{\kappab}_{mn} \widehat{\varphi}) + \epsilon_{mnr}\theta_{mnr} (\varphi \etab - \widehat{\etab} \widehat{\varphi}) \nn \\
&&~~~~- \psi_m (\widehat{\phi} \widehat{\psib}_m - \psib_m \phi) - \theta_{mnr} (\widehat{\phi} \widehat{\thetab}_{mnr} - \thetab_{mnr} \phi) \nn \\
&&~~~~- \epsilon_{mnr} \eta (\widehat{\varphib} \widehat{\thetab}_{mnr} - \thetab_{mnr} \varphib) - \eta (\phib \etab - \widehat{\etab} \widehat{\phib}) \nn \\
&&~~~~- \hf \chi_{mn} (\phib  \kappab_{mn} - \widehat{\kappab}_{mn} \widehat{\phib}) + \epsilon_{mnr} \chi_{mn} (\widehat{\varphib} \widehat{\psib}_r - \psib_r \varphib) \nn \\
&&~~~~+ \hf \Big(\widehat{\varphib} \widehat{\varphi} - \varphi \varphib+ \phib \phi - \widehat{\phi} \widehat{\phib} \Big)^2 - 2 (\widehat{\varphib} \widehat{\phib} - \phib \varphib)(\varphi \phi - \widehat{\phi} \widehat{\varphi}) \Big]
\eea
and
\bea
S^{\rm matter}_{(\Boxb, \Box)}  &=& \frac{1}{g_3^2} \int d^3x \Tr \Big[2 \varphib \cDb_m \cD_m \varphi + 2 \widehat{\phib} \cDb_m \cD_m \widehat{\phi} + [\cDb_m, \cD_m] \Big(\varphib \varphi - \widehat{\varphi} \widehat{\varphib} + \widehat{\phib} \widehat{\phi}  - \phi \phib \Big) \nn \\
&&~~~~+ \widehat{\etab} \cD_m \widehat{\psib}_m - \kappab_{np} \cDb_p \psib_n + \hf \widehat{\thetab}_{mnr} \cD_m \widehat{\kappab}_{nr} \nn \\
&&~~~~- \hf \epsilon_{mnr}\widehat{\psi}_r (\widehat{\varphi} \widehat{\kappab}_{mn} - \kappab_{mn} \varphi) + \epsilon_{mnr} \widehat{\theta}_{mnr} (\widehat{\varphi} \widehat{\etab} - \etab \varphi) \nn \\
&&~~~~- \widehat{\psi}_m (\phi \psib_m - \widehat{\psib}_m \widehat{\phi}) - \widehat{\theta}_{mnr} (\phi \thetab_{mnr} - \widehat{\thetab}_{mnr} \widehat{\phi}) \nn \\
&&~~~~- \epsilon_{mnr} \widehat{\eta} (\varphib \thetab_{mnr} - \widehat{\thetab}_{mnr} \widehat{\varphib}) - \widehat{\eta} (\widehat{\phib} \widehat{\etab} - \etab \phib) \nn \\
&&~~~~- \hf \widehat{\chi}_{mn} (\widehat{\phib}  \widehat{\kappab}_{mn} - \kappab_{mn} \phib) + \epsilon_{mnr} \widehat{\chi}_{mn} (\varphib \psib_r - \widehat{\psib}_r \widehat{\varphib}) \nn \\
&&~~~~+ \hf \Big(\varphib \varphi - \widehat{\varphi} \widehat{\varphib} + \widehat{\phib} \widehat{\phi} - \phi \phib \Big)^2 - 2 (\varphib \phib - \widehat{\phib} \widehat{\varphib})(\widehat{\varphi} \widehat{\phi} - \phi \varphi) \Big],
\eea
with the field content $\{\cA_m$, $\cAb_m$, $\eta$, $\psi_m$, $\chi_{mn}$, $\theta_{mnr}\}$, $\{\widehat{\cA}_m$, $\widehat{\cAb}_m$, $\widehat{\eta}$, $\widehat{\psi}_m$, $\widehat{\chi}_{mn}$, $\widehat{\theta}_{mnr}\}$, $\{ \widehat{\phi}, \phib, \varphi, \widehat{\varphib}, \widehat{\etab}, \psib_m, \widehat{\kappab}_{mn}, \thetab_{mnr} \}$ and $\{\phi, \widehat{\phib}, \widehat{\varphi}, \varphib, \etab, \widehat{\psib}_m, \kappab_{mn}, \widehat{\thetab}_{mnr}\}$ transforming respectively as $({\bf adj}, {\bf 1})$, $({\bf 1}, {\bf adj})$, $(\Box, \Boxb)$ and $(\Boxb, \Box)$ under $U(N_1) \times U(N_2)$.

\section{Lattice formulation of supersymmetric quiver theories}
\label{sec:lattice_susy_quiver_theories}

The twisted supersymmetric gauge theories and their quiver cousins described in the previous sections can be discretized on a Euclidean spacetime lattice in a straightforward manner. We use the method of geometric discretization developed in refs. \cite{Catterall:2007kn, Damgaard:2007be, Damgaard:2008pa}. The continuum complex gauge fields $\cA_m(x)$ at every spacetime point are mapped to appropriate complexified Wilson links $\cU_m(\vn)$. These complex link fields are taken to be associated with unit length vectors in the coordinate directions $\hatbnu_m$ from the site denoted by the integer vector $\vn$ on a hypercubic lattice. Supersymmetric invariance then implies that the components of the 1-form fermion field $\psi_m(\vn)$ live on the same oriented links as that of their bosonic superpartners, $\cU_m(\vn)$, running from $\vn \to \vn + \hatbnu_m$. The scalar fermion $\eta(\vn)$ is associated with the site $\vn$ of the lattice. The components of the 2-form field $\chi_{mn}(\vn)$ are placed on a set of diagonal face links, running from $\vn + \hatbnu_m + \hatbnu_n \to \vn$. The 3-form field $\theta_{mnr}(\vn)$ is placed on the body diagonal, along the direction $\vn \to \vn + \hatbnu_m + \hatbnu_n + \hatbnu_r$.

In general, the prescription for geometric discretization of topologically twisted field theories is the following: lattice variables $\cU_m(\vn)$, $\cUb_m(\vn)$, $\{f^{(+)}_{m_1 \cdots m_p}(\vn) \}$, $\{f^{(-)}_{m_1 \cdots m_p}(\vn) \}$ should be associated with the links $(\vn, \vn + \hatbnu_m)$, $(\vn + \hatbnu_m, \vn)$, $(\vn, \vn + \hatbnu_{m_1} + \cdots + \hatbnu_{m_p})$ and $(\vn + \hatbnu_{m_1} + \cdots + \hatbnu_{m_p}, \vn)$, respectively. A site variable $f(\vn)$ should be associated with a degenerate link $(\vn, \vn)$. 

We can write down the gauge transformation rules for the adjoint lattice fields respecting the p-cell and orientation assignments on the lattice. For $G(\vn) \in U(N)$, we have the following gauge transformation prescriptions \cite{Aratyn:1984bd, Damgaard:2008pa}
\bea
\label{eq:gauge_lattice_1}
\cU_m(\vn) &\rightarrow& G(\vn) \cU_m(\vn) G^{\dagger}(\vn + \hatbnu_m), \\ 
\cUb_m(\vn) &\rightarrow& G(\vn + \hatbnu_m) \cUb_m(\vn)G^{\dagger}(\vn), \\ 
\{f^{(+)}_{m_1 \cdots m_p}(\vn) \} &\rightarrow& G(\vn) \{f^{(+)}_{m_1 \cdots m_p}(\vn) \} G^{\dagger}(\vn + \hatbnu_{m_1} + \cdots + \hatbnu_{m_p}), \\ 
\label{eq:gauge_lattice_2}
\{f^{(-)}_{m_1 \cdots m_p}(\vn) \} &\rightarrow& G(\vn + \hatbnu_{m_1} + \cdots + \hatbnu_{m_p} ) \{f^{(-)}_{m_1 \cdots m_p}(\vn) \} G^{\dagger}(\vn).
\eea

We need to describe how continuum covariant derivatives are to be replaced by covariant difference operators. The covariant derivatives $\cD_m$ ($\cDb_m$) in the continuum become forward and backward covariant differences $\cD^{(+)}_m$ ($\cDb^{(+)}_m$) and $\cD^{(-)}_m$ ($\cDb^{(-)}_m$), respectively on the lattice. The forward covariant difference operators act on the lattice fields $f^{(\pm)}_{m_1 \cdots m_p}(\vn)$ in the following way
\bea
\label{eq:cov_diff_adj_1}
&&\cD_n^{(+)}f^{(+)}_{m_1 \cdots m_p}(\vn) \equiv \cU_n(\vn)f^{(+)}_{m_1 \cdots m_p}(\vn + \hatbnu_n) - f^{(+)}_{m_1 \cdots m_p}(\vn) \cU_n(\vn + \hatbnu),~~~~~~~~ \\
&&\cD_n^{(+)}f^{(-)}_{m_1 \cdots m_p}(\vn) \equiv \cU_n(\vn + \hatbnu)f^{(-)}_{m_1 \cdots m_p}(\vn + \hatbnu_n)-f^{(-)}_{m_1 \cdots m_p}(\vn) \cU_n(\vn),~~~~~~~~ \\
&&\cDb_n^{(+)}f^{(+)}_{m_1 \cdots m_p}(\vn) \equiv f^{(+)}_{m_1 \cdots m_p}(\vn + \hatbnu_n)\cUb_n(\vn + \hatbnu) - \cUb_n(\vn)f^{(+)}_{m_1 \cdots m_p}(\vn),~~~~~~~~ \\
&&\cDb_n^{(+)}f^{(-)}_{m_1 \cdots m_p}(\vn) \equiv f^{(-)}_{m_1 \cdots m_p}(\vn + \hatbnu_n)\cUb_n(\vn) - \cUb_n(\vn + \hatbnu)f^{(-)}_{m_1 \cdots m_p}(\vn),~~~~~~~~
\eea
where we have defined $\hatbnu = \sum_{i=1}^p \hatbnu_{m_i}$. 

The action of the backward covariant difference operators on the lattice fields is given by 
\bea
\cD_n^{(-)}f^{(\pm)}_{m_1 \cdots m_p}(\vn) &\equiv& \cD_n^{(+)}f^{(\pm)}_{m_1 \cdots m_p}(\vn - \hatbnu_n), \\
\label{eq:cov_diff_adj_2}
\cDb_n^{(-)}f^{(\pm)}_{m_1 \cdots m_p}(\vn) &\equiv& \cDb_n^{(+)}f^{(\pm)}_{m_1 \cdots m_p}(\vn - \hatbnu_n).
\eea

These expressions are determined by the two requirements that they reduce to the corresponding continuum results for the adjoint covariant derivative in the naive continuum limit and that they transform under gauge transformations like the corresponding lattice link field carrying the same indices. As a result, the terms in the lattice action correspond to gauge-invariant closed loops. 

The lattice field strength is given by the expression $\cF_{mn}(\vn) = \cD^{(+)}_m \cU_n(\vn)$. We see that it is automatically antisymmetric in its indices and also it transforms like a lattice 2-form.

We also need to define the action of the covariant difference operators on the lattice fields transforming in the bi-fundamental representations. In the lattice constructions of the quiver gauge theories with gauge group $U(N_1) \times U(N_2)$ we have two spacetime lattices with same dimensionalities, corresponding to the two nodes of the quiver, which we label as the $N_1$-lattice and $N_2$-lattice, respectively. We denote the position on the $N_1$-lattice by an integer valued vector $\vn$ while the same position on the $N_2$-lattice is denoted by the vector $\uvn$. The fields on the $N_1$-lattice transform as $({\bf adj}, {\bf 1})$ while those on the $N_2$-lattice transform as $({\bf 1}, {\bf adj})$ under the gauge group $U(N_1) \times U(N_2)$. The action of the forward and backward covariant difference operators on adjoint fields living on $N_1$- and $N_2$-lattices is summarized in eqs. (\ref{eq:cov_diff_adj_1}) - (\ref{eq:cov_diff_adj_2}).  

The bi-fundamental matter fields of the lattice quiver theory live on the links connecting the $N_1$- and $N_2$-lattice spacetimes. They transform in the bi-fundamental representations of $U(N_1) \times U(N_2)$. We have the following set of rules for the action of the covariant derivatives on bi-fundamental fields. 

For lattice variables in the representation $(\Box, \Boxb)$ the covariant forward difference operators act on them the following way
\bea
&&\cD_n^{(+)} f^{(+)}_{m_1 \cdots m_p}(\vn, \uvn) \equiv \cU_n(\vn) f^{(+)}_{m_1 \cdots m_p}(\vn + \hatbnu_n, \uvn + \uhatbnu_n) - f^{(+)}_{m_1 \cdots m_p}(\vn, \uvn) \widehat{\cU}_n(\uvn),~~~~~~~~ \\
&&\cD_n^{(+)} f^{(-)}_{m_1 \cdots m_p}(\vn, \uvn) \equiv \cU_n(\vn + \hatbnu) f^{(-)}_{m_1 \cdots m_p}(\vn + \hatbnu_n, \uvn + \uhatbnu_n) - f^{(-)}_{m_1 \cdots m_p}(\vn, \uvn) \widehat{\cU}_n(\uvn),~~~~~~~~ \\
&&\cDb_n^{(+)} f^{(+)}_{m_1 \cdots m_p}(\vn, \uvn) \equiv f^{(+)}_{m_1 \cdots m_p}(\vn + \hatbnu_n, \uvn + \uhatbnu_n) \widehat{\cUb}_n(\uvn) - \cUb_n(\vn) f^{(+)}_{m_1 \cdots m_p}(\vn, \uvn),~~~~~~~~ \\
&&\cDb_n^{(+)} f^{(-)}_{m_1 \cdots m_p}(\vn, \uvn) \equiv f^{(-)}_{m_1 \cdots m_p}(\vn + \hatbnu_n, \uvn + \uhatbnu_n) \widehat{\cUb}_n(\uvn) - \cUb_n(\vn + \hatbnu)f^{(-)}_{m_1 \cdots m_p}(\vn, \uvn),
\eea
while the covariant backward difference operators act on the fields according to the rules
\bea
\cD_n^{(-)} f^{(\pm)}_{m_1 \cdots m_p}(\vn, \uvn) &\equiv& \cD_n^{(+)} f^{(\pm)}_{m_1 \cdots m_p}(\vn - \hatbnu_n, \uvn - \uhatbnu_n), \\
\cDb_n^{(-)} f^{(\pm)}_{m_1 \cdots m_p}(\vn, \uvn) &\equiv& \cDb_n^{(+)} f^{(\pm)}_{m_1 \cdots m_p}(\vn - \hatbnu_n, \uvn - \uhatbnu_n).
\eea

For lattice variables in the representation $(\Boxb, \Box)$ we have the following set of rules for the covariant difference operators
\bea
&&\cD_n^{(+)} f^{(+)}_{m_1 \cdots m_p}(\uvn, \vn) \equiv \widehat{\cU}_n(\uvn) f^{(+)}_{m_1 \cdots m_p}(\uvn + \uhatbnu_n, \vn + \hatbnu_n) - f^{(+)}_{m_1 \cdots m_p}(\uvn, \vn) \cU_n(\vn),~~~~~~~~ \\
&&\cD_n^{(+)} f^{(-)}_{m_1 \cdots m_p}(\uvn, \vn) \equiv \widehat{\cU}_n(\uvn + \uhatbnu) f^{(-)}_{m_1 \cdots m_p}(\uvn + \uhatbnu_n, \vn + \hatbnu_n) - f^{(-)}_{m_1 \cdots m_p}(\uvn, \vn) \cU_n(\vn),~~~~~~~~ \\
&&\cDb_n^{(+)} f^{(+)}_{m_1 \cdots m_p}(\uvn, \vn) \equiv f^{(+)}_{m_1 \cdots m_p}(\uvn + \uhatbnu_n, \vn + \hatbnu_n) \cUb_n(\vn) - \widehat{\cUb}_n(\uvn) f^{(+)}_{m_1 \cdots m_p}(\uvn, \vn),~~~~~~~~ \\
&&\cDb_n^{(+)} f^{(-)}_{m_1 \cdots m_p}(\uvn, \vn) \equiv f^{(-)}_{m_1 \cdots m_p}(\uvn + \uhatbnu_n, \vn + \hatbnu_n)\cUb_n(\vn) - \widehat{\cUb}_n(\uvn + \uhatbnu)f^{(-)}_{m_1 \cdots m_p}(\uvn, \vn),
\eea
and
\bea
\cD_n^{(-)} f^{(\pm)}_{m_1 \cdots m_p}(\uvn, \vn) &\equiv& \cD_n^{(+)} f^{(\pm)}_{m_1 \cdots m_p}(\uvn - \uhatbnu_n, \vn - \hatbnu_n), \\
\cDb_n^{(-)} f^{(\pm)}_{m_1 \cdots m_p}(\uvn, \vn) &\equiv& \cDb_n^{(+)} f^{(\pm)}_{m_1 \cdots m_p}(\uvn - \uhatbnu_n, \vn - \hatbnu_n).
\eea

We also note that the method of geometric discretization maps the continuum fields on to the lattice one-to-one and thus the lattice theories constructed this way are free from fermion doubling problem \cite{Rabin:1981qj, Becher:1982ud, Banks:1982iq, Aratyn:1984bd}. Now that we have the rules for implementing quiver gauge theories on the lattice we move on to the lattice constructions of quiver gauge theories discussed in the previous sections.

\subsection{Two-dimensional $Q=4$ lattice quiver gauge theory}
\label{sec:2d_q4_lattice_quiver}

\begin{figure}
\begin{center}
\includegraphics[width=0.45\textwidth]{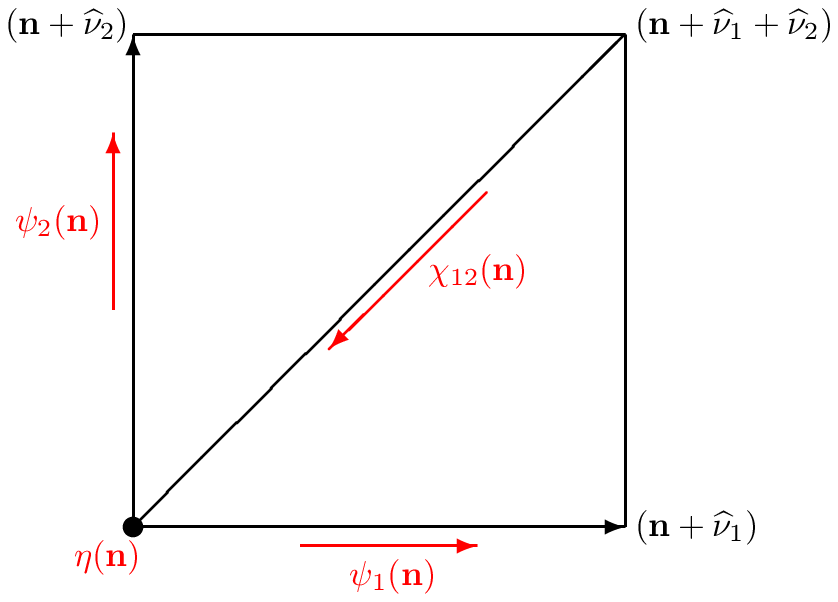}
\end{center}
\caption{\label{fig:2dq4latt}The unit cell of the two-dimensional $Q=4$ lattice SYM with orientation assignments for the twisted fermions. The complexified bosons $\cU_a$ follow the same orientations and link assignments as that of their superpartners $\psi_a$.}
\end{figure}

The two-dimensional $Q=4$ lattice quiver gauge theory with gauge group $U(N_1) \times U(N_2)$ contains two two-dimensional lattice spacetimes ($N_1$- and $N_2$-lattice) corresponding to each node of the quiver. The unit cell is a square lattice. For the $N_1$-lattice, the adjoint fermion fields, $\eta(\vn)$, $\psi_a(\vn)$ and $\chi_{ab}(\vn)$, with gauge group $U(N_1)$, live on the site, edge and diagonal link, respectively, of the unit cell. The complexified Wilson links $\cU_a(\vn)$ and $\cUb_a(\vn)$ also live on the edges of the unit cell. The placement and orientations of the twisted fields on the lattice respect the scalar supersymmetry and gauge symmetry of the lattice theory. The unit cell of the two dimensional lattice theory is given in figure \ref{fig:2dq4latt}. The other set of adjoint fields, decorated with hats, live on the $N_2$-lattice and transform as adjoints under the gauge group $U(N_2)$. 

The action of the two-dimensional $Q=4$ lattice quiver gauge theory has the form
\beq
S = S^{\rm SYM}_{({\bf adj}, {\bf 1})} + S^{\rm SYM}_{({\bf 1}, {\bf adj})} + S^{\rm matter}_{(\Box, \Boxb)} + S^{\rm matter}_{(\Boxb, \Box)},
\eeq
where
\bea
S^{\rm SYM}_{({\bf adj}, {\bf 1})} &=& \frac{1}{g_2^2}\sum_{\vn} \Tr \Big\{\Big(\cUb_b(\vn + \hatbnu_a)\cUb_a(\vn) - \cUb_a(\vn + \hatbnu_b)\cUb_b(\vn)\Big)\nn \\
&&~~~~\times \Big(\cU_a(\vn)\cU_b(\vn + \hatbnu_a) - \cU_b(\vn)\cU_a(\vn + \hatbnu_b)\Big) \nn \\
&&~~~~+ \hf \Big(\cU_a(\vn)\cUb_a(\vn) - \cUb_a(\vn - \hatbnu_a)\cU_a(\vn - \hatbnu_a)\Big)^2 \nn \\
&&~~~~+\hf(\delta_{aq}\delta_{br} - \delta_{ar}\delta_{bq}) \chi_{ab}(\vn) \Big(\cU_q(\vn)\psi_r(\vn + \hatbnu_q) - \psi_r(\vn)\cU_q(\vn + \hatbnu_r)\Big) \nn \\
&&~~~~+ \eta(\vn)\Big(\psi_a(\vn)\cUb_a(\vn) - \cUb_a(\vn - \hatbnu_a)\psi_a(\vn - \hatbnu_a)\Big)\Big\},
\eea

\bea
S^{\rm SYM}_{({\bf 1}, {\bf adj})} &=& \frac{1}{g_2^2}\sum_{\uvn} \Tr \Big\{\Big(\widehat{\cUb}_b(\uvn + \uhatbnu_a)\widehat{\cUb}_a(\uvn) - \widehat{\cUb}_a(\uvn + \uhatbnu_b)\widehat{\cUb}_b(\uvn)\Big)\nn \\
&&~~~~\times \Big(\widehat{\cU}_a(\uvn)\widehat{\cU}_b(\uvn + \uhatbnu_a) - \widehat{\cU}_b(\uvn)\widehat{\cU}_a(\uvn + \uhatbnu_b)\Big) \nn \\
&&~~~~+ \hf \Big(\widehat{\cU}_a(\uvn)\widehat{\cUb}_a(\uvn) - \widehat{\cUb}_a(\uvn - \uhatbnu_a)\widehat{\cU}_a(\uvn - \uhatbnu_a)\Big)^2 \nn \\
&&~~~~+\hf(\delta_{aq}\delta_{br} - \delta_{ar}\delta_{bq}) \widehat{\chi}_{ab}(\uvn) \Big(\widehat{\cU}_q(\uvn)\widehat{\psi}_r(\uvn + \uhatbnu_q) - \widehat{\psi}_r(\uvn)\widehat{\cU}_q(\uvn + \uhatbnu_r)\Big) \nn \\
&&~~~~+ \widehat{\eta}(\uvn)\Big(\widehat{\psi}_a(\uvn)\widehat{\cUb}_a(\uvn) - \widehat{\cUb}_a(\uvn - \uhatbnu_a)\widehat{\psi}_a(\uvn - \uhatbnu_a)\Big)\Big\},
\eea

\begin{figure}
\begin{center}
\includegraphics[width=0.6\textwidth]{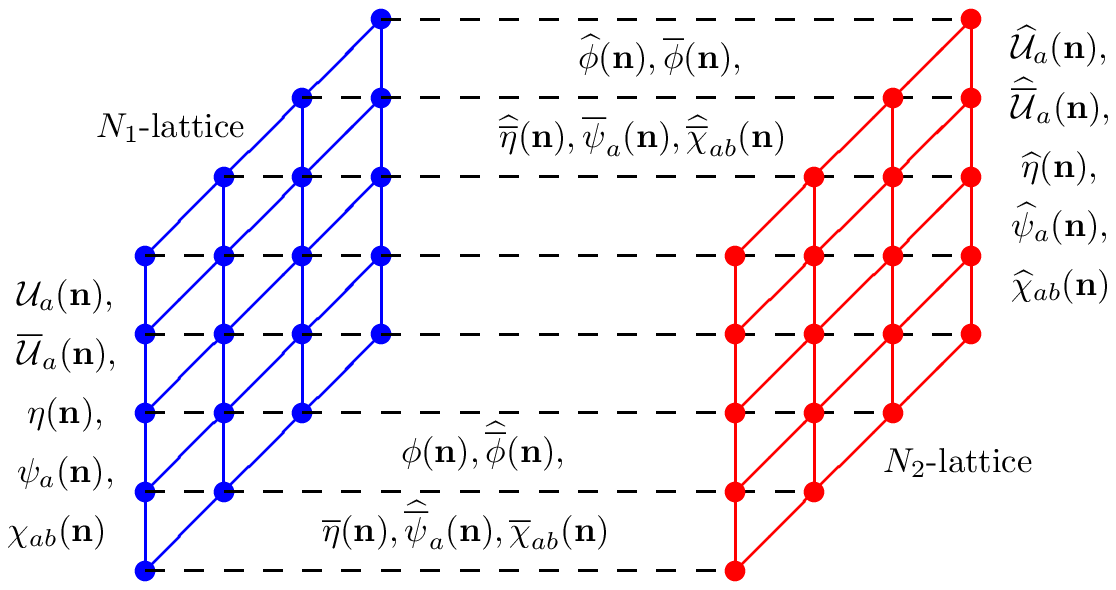}
\end{center}
\caption{\label{fig:2dq4_quiver_latt}Schematic sketch of the lattice construction of two-dimensional $Q=4$ quiver gauge theory. The lattice variables $\{\cU_a(\vn)$, $\cUb_a(\vn)$, $\eta(\vn)$, $\psi_a(\vn)$, $\chi_{ab}(\vn)\}$ live on the two-dimensional $N_1$-lattice spacetime and $\{\widehat{\cU}_a(\vn)$, $\widehat{\cUb}_a(\vn)$, $\widehat{\eta}(\vn)$, $\widehat{\psi}_a(\vn)$, $\widehat{\chi}_{ab}(\vn)\}$ live on the two-dimensional $N_2$-lattice spacetime. The matter fields $\{\phi(\vn)$, $\widehat{\phib}(\vn)$, $\etab(\vn)$, $\widehat{\psib}_a(\vn)$, $\chib_{ab}(\vn)\}$ and  $\{\widehat{\phi}(\vn)$, $\phib(\vn)$, $\widehat{\etab}(\vn)$, $\psib_a(\vn)$, $\widehat{\chib}_{ab}(\vn)\}$ live on the links connecting the two lattice spactimes.}
\end{figure}

\bea
S^{\rm matter}_{(\Box, \Boxb)} &=& \frac{1}{g_2^2}\sum_{\vn} \Tr \Big\{2 \widehat{\phib}(\vn, \uvn) \cDb^{(-)}_a \cD^{(+)}_a \widehat{\phi}(\uvn, \vn) \nn \\
&&~~~~+ \Big(\cDb^{(-)}_a \cU_a(\vn)\Big) \Big(\widehat{\phib}(\vn, \uvn) \widehat{\phi}(\uvn, \vn) - \phi(\vn, \uvn) \phib(\uvn, \vn)\Big) \nn \\
&&~~~~- \chib_{ab}(\vn, \uvn) \cDb^{(+)}_b \psib_a(\uvn, \vn) + \widehat{\psib}_a(\vn, \uvn) \cD^{(+)}_a \widehat{\etab}(\uvn, \vn) \nn \\
&&~~~~- \psi_a(\vn, \vn) \Big(\phi(\vn, \uvn) \psib_a(\uvn, \vn) - \widehat{\psib}_a(\vn, \uvn) \widehat{\phi}(\uvn, \vn)\Big) \nn \\
&&~~~~- \eta(\vn, \vn) \Big(\widehat{\phib}(\vn, \uvn) \widehat{\etab}(\uvn, \vn) - \etab(\vn, \uvn) \phib(\uvn, \vn)\Big) \nn \\
&&~~~~- \hf \chi_{ab}(\vn, \vn) \Big(\widehat{\phib}(\vn, \uvn) \widehat{\chib}_{ab}(\uvn, \vn) - \chib_{ab}(\vn, \uvn) \phib(\uvn, \vn)\Big) \nn \\
&&~~~~+ \hf (\widehat{\phib}(\vn, \uvn) \widehat{\phi}(\uvn, \vn) - \phi(\vn, \uvn) \phib(\uvn, \vn))^2 \Big\}
\eea
and
\bea
S^{\rm matter}_{(\Boxb, \Box)} &=& \frac{1}{g_2^2}\sum_{\uvn} \Tr \Big\{2 \phib(\uvn, \vn) \cDb^{(-)}_a \cD^{(+)}_a \phi(\vn, \uvn) \nn \\
&&~~~~+ \Big(\cDb^{(-)}_a \cU_a(\uvn)\Big) \Big(\phib(\uvn, \vn) \phi(\vn, \uvn) - \widehat{\phi}(\uvn, \vn) \widehat{\phib}(\vn, \uvn)\Big) \nn \\
&&~~~~- \widehat{\chib}_{ab}(\uvn, \vn) \cDb^{(+)}_b \widehat{\psib}_a(\vn, \uvn) + \psib_a(\uvn, \vn) \cD^{(+)}_a \etab(\vn, \uvn) \nn \\
&&~~~~- \widehat{\psi}_a(\uvn, \uvn) \Big(\widehat{\phi}(\uvn, \vn) \widehat{\psib}_a(\vn, \uvn) - \psib_a (\uvn, \vn)\phi(\vn, \uvn)\Big) \nn \\
&&~~~~- \widehat{\eta}(\uvn, \uvn) \Big(\phib(\uvn, \vn) \etab(\vn, \uvn) - \widehat{\etab}(\uvn, \vn) \widehat{\phib}(\vn, \uvn)\Big) \nn \\
&&~~~~- \hf \widehat{\chi}_{ab}(\uvn, \uvn) \Big(\phib(\uvn, \vn) \chib_{ab}(\vn, \uvn) - \widehat{\chib}_{ab}(\uvn, \vn) \widehat{\phib}(\vn, \uvn)\Big) \nn \\
&&~~~~+ \hf (\phib(\uvn, \vn) \phi(\vn, \uvn) - \widehat{\phi}(\uvn, \vn) \widehat{\phib}(\vn, \uvn))^2 \Big\}.
\eea

In figure \ref{fig:2dq4_quiver_latt} we provide a schematic sketch of the two-dimensional $Q=4$ lattice quiver gauge theory with the placement of adjoint and bi-fundamental fields. 

\subsection{Two-dimensional $Q=8$ lattice quiver gauge theory}
\label{sec:2d_q8_lattice_quiver}

\begin{figure}
\begin{center}
\includegraphics[width=0.45\textwidth]{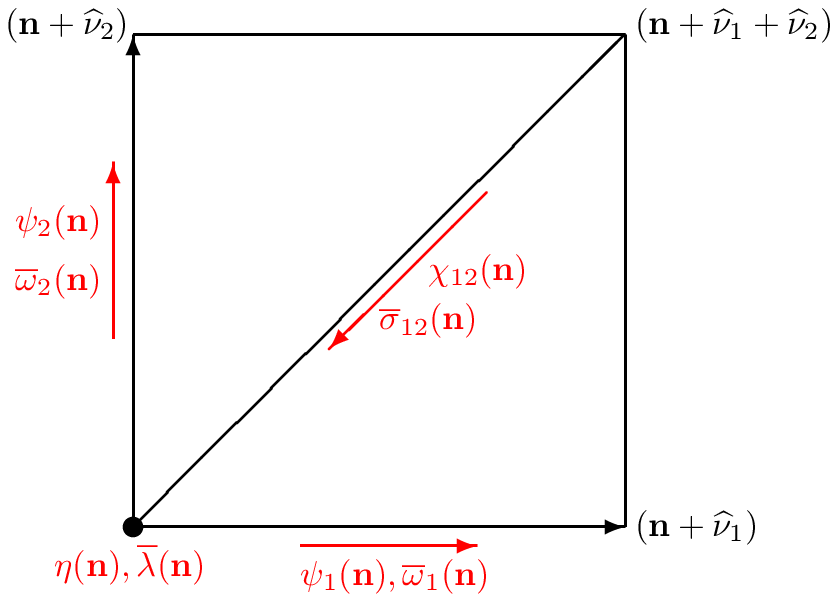}
\end{center}
\caption{\label{fig:2dq8latt}The unit cell of the two-dimensional $Q=8$ lattice SYM with orientation assignments for twisted fermions. The complexified bosons $\cU_a$ follow the same orientations and link assignments as that of their superpartners $\psi_a$ and $\omegab_a$. The scalar fields $\pi(\vn)$ and $\pib(\vn)$ are placed at the origin of the unit cell along with $\eta(\vn)$ and $\lambdab(\vn)$.}
\end{figure}

The two-dimensional $Q=8$ lattice quiver gauge theory with gauge group $U(N_1) \times U(N_2)$ also contains two two-dimensional lattice spacetimes ($N_1$- and $N_2$-lattice) corresponding to each node of the quiver. The unit cell is a square lattice and the twisted fermions are distributed on the unit cell in multiplets of two. For the $N_1$-lattice, the adjoint fermion fields, $\eta(\vn)$ and $\lambdab(\vn)$ live on a site, $\psi_a(\vn)$ and $\omegab_a(\vn)$ live on a link and $\chi_{ab}(\vn)$ and $\sigmab_{ab}(\vn)$ live on a diagonal link of the unit cell. They all transform in the adjoint representation of the gauge group $U(N_1)$. The complexified Wilson links $\cU_a(\vn)$ and $\cUb_a(\vn)$ reside on the edges of the unit cell. The unit cell of the two dimensional $Q=8$ lattice theory is given in figure \ref{fig:2dq8latt}. The set of adjoint fields that are decorated with hats live on the $N_2$-lattice spacetime and transform as adjoints under the gauge group $U(N_2)$. 

The action of the two-dimensional $Q=8$ lattice quiver gauge theory has the following form
\beq
S = S^{\rm SYM}_{({\bf adj}, {\bf 1})} + S^{\rm SYM}_{({\bf 1}, {\bf adj})} + S^{\rm matter}_{(\Box, \Boxb)} + S^{\rm matter}_{(\Boxb, \Box)},
\eeq
where
\bea
S^{\rm SYM}_{({\bf adj}, {\bf 1})} &=& \frac{1}{g_2^2} \sum_{\vn} \Tr \Big\{\Big(\cUb_b(\vn + \hatbnu_a)\cUb_a(\vn) - \cUb_a(\vn + \hatbnu_b)\cUb_b(\vn)\Big)\nn \\
&&~~~~\times \Big(\cU_a(\vn)\cU_b(\vn + \hatbnu_a) - \cU_b(\vn)\cU_a(\vn + \hatbnu_b)\Big) \nn \\
&&~~~~+ \hf \Big(\cU_a(\vn)\cUb_a(\vn) - \cUb_a(\vn - \hatbnu_a)\cU_a(\vn - \hatbnu_a)\Big)^2 \nn \\
&&~~~~+ 2 \pib(\vn) \cDb^{(-)}_a \cD^{(+)}_a \pi(\vn) + \Big(\cDb^{(-)}_a \cU_a(\vn)\Big)[\pib(\vn), \pi(\vn)] \nn \\
&&~~~~+ \hf [\pib(\vn), \pi(\vn)]^2 - \eta(\vn) \cDb^{(-)}_a \psi_a(\vn) - \chi_{ab}(\vn) \cD^{(+)}_a \psi_b(\vn) \nn \\
&&~~~~+ \lambdab(\vn) \cD^{(-)}_a \omegab_a(\vn) - \sigmab_{ab}(\vn)\cDb^{(+)}_b \omegab_a(\vn) - \eta(\vn) [\pib(\vn), \lambdab(\vn)] \nn \\
&&~~~~- \psi_a(\vn) [\pi(\vn), \omegab_a(\vn)] - \hf \chi_{ab}(\vn) [\pib(\vn),\sigmab_{ab}(\vn)] \Big\},
\eea

\bea
S^{\rm SYM}_{({\bf 1}, {\bf adj})} &=& \frac{1}{g_2^2} \sum_{\uvn} \Tr \Big\{\Big(\widehat{\cUb}_b(\uvn + \uhatbnu_a)\widehat{\cUb}_a(\uvn) - \widehat{\cUb}_a(\uvn + \uhatbnu_b)\widehat{\cUb}_b(\uvn)\Big) \nn \\
&&~~~~\times \Big(\widehat{\cU}_a(\uvn)\widehat{\cU}_b(\uvn + \uhatbnu_a) - \widehat{\cU}_b(\uvn)\widehat{\cU}_a(\uvn + \uhatbnu_b)\Big) \nn \\
&&~~~~+ \hf \Big(\widehat{\cU}_a(\uvn)\widehat{\cUb}_a(\uvn) - \widehat{\cUb}_a(\uvn - \uhatbnu_a)\widehat{\cU}_a(\uvn - \uhatbnu_a)\Big)^2 \nn \\
&&~~~~+ 2 \widehat{\pib}(\uvn) \widehat{\cDb}^{(-)}_a \widehat{\cD}^{(+)}_a \widehat{\pi}(\uvn) + \Big(\widehat{\cDb}^{(-)}_a \widehat{\cU}_a(\uvn)\Big) [\widehat{\pib}(\uvn), \widehat{\pi}(\uvn)] \nn \\
&&~~~~+ \hf [\widehat{\pib}(\uvn), \widehat{\pi}(\uvn)]^2 - \widehat{\eta}(\uvn) \widehat{\cDb}^{(-)}_a \widehat{\psi}_a(\uvn) - \widehat{\chi}_{ab}(\uvn) \widehat{\cD}^{(+)}_a \widehat{\psi}_b(\uvn) \nn \\
&&~~~~+ \widehat{\lambdab}(\uvn) \widehat{\cD}^{(-)}_a \widehat{\omegab}_a(\uvn) - \widehat{\sigmab}_{ab}(\uvn) \widehat{\cDb}^{(+)}_b \widehat{\omegab}_a(\uvn) - \widehat{\eta}(\uvn) [\widehat{\pib}(\uvn), \widehat{\lambdab}(\uvn)] \nn \\ 
&&~~~~- \widehat{\psi}_a(\uvn) [\widehat{\pi}(\uvn), \widehat{\omegab}_a(\uvn)] - \hf \widehat{\chi}_{ab}(\uvn) [\widehat{\pib}(\uvn), \widehat{\sigmab}_{ab}(\uvn)] \Big\},
\eea

\bea
S^{\rm matter}_{(\Box, \Boxb)} &=& \frac{1}{g_2^2} \sum_{\vn} \Tr \Big\{2 \widehat{\varphib}(\vn, \uvn) \cDb^{(-)}_a\cD^{(+)}_a \widehat{\varphi}(\uvn, \vn) + 2 \phib(\vn, \uvn) \cDb^{(-)}_a\cD^{(+)}_a \phi(\uvn, \vn) \nn \\
&&+ \Big(\cDb^{(-)}_a \cU_a(\vn)\Big) \Big(\widehat{\varphib}(\vn, \uvn) \widehat{\varphi}(\uvn, \vn) - \varphi(\vn, \uvn) \varphib(\uvn, \vn) + \phib(\vn, \uvn) \phi(\uvn, \vn) \nn \\
&&- \widehat{\phi}(\vn, \uvn) \widehat{\phib}(\uvn, \vn)\Big) + \etab(\vn, \uvn) \cD^{(-)}_a \psib_a(\uvn, \vn) - \widehat{\rhob}(\vn, \uvn) \cDb^{(-)}_a \widehat{\taub}_a(\uvn, \vn) \nn \\
&&- \widehat{\kappab}_{ab}(\vn, \uvn) \cDb^{(+)}_a \widehat{\psib}_b(\uvn, \vn) - 2 \xib_{ab}(\vn, \uvn) \cD^{(+)}_a \taub_b(\uvn, \vn) \nn \\
&&+ \rhob(\vn, \uvn) \Big(\widehat{\pi}(\uvn, \uvn) \etab(\uvn, \vn) - \etab(\uvn, \vn) \pi(\vn, \vn)\Big) + \widehat{\taub}_a(\vn, \uvn) \Big(\widehat{\pib}(\uvn, \uvn) \widehat{\psib}_a(\uvn, \vn) \nn \\
&&- \widehat{\psib}_a(\uvn, \vn) \pib(\vn, \vn)\Big) + \hf \xib_{ab}(\vn, \uvn) \Big(\widehat{\pi}(\uvn, \uvn) \kappab_{ab}(\uvn, \vn) - \kappab_{ab}(\uvn, \vn) \pi(\vn, \vn)\Big) \nn \\
&&- \eta(\vn, \vn) \Big(\phib(\vn, \uvn) \etab(\uvn, \vn) - \widehat{\etab}(\vn, \uvn) \widehat{\phib}(\uvn, \vn)\Big) - 3\epsilon_{ab} \eta(\vn, \vn) \Big(\widehat{\varphib}(\vn, \uvn) \widehat{\xib}_{ab}(\uvn, \vn) \nn \\
&&- \xib_{ab}(\vn, \uvn) \varphib(\uvn, \vn)\Big) + \frac{3}{2} \epsilon_{ab} \sigmab_{ab}(\vn, \vn) \Big(\varphi(\vn, \uvn) \etab(\uvn, \vn) - \widehat{\etab}(\vn, \uvn) \widehat{\varphi}(\uvn, \vn)\Big) \nn \\
&&- \psi_a(\vn, \vn) \Big(\widehat{\phi}(\vn, \uvn) \widehat{\psib}_a(\uvn, \vn) - \psib_a(\vn, \uvn) \phi(\uvn, \vn)\Big) - \lambdab(\vn, \vn) \Big(\widehat{\phi}(\vn, \uvn) \widehat{\rhob}(\uvn, \vn) \nn \\
&&- \rhob(\vn, \uvn) \phi(\uvn, \vn)\Big) - \hf \epsilon_{ab} \lambdab(\vn, \vn) \Big(\varphi(\vn, \uvn) \kappab_{ab}(\uvn, \vn) - \widehat{\kappab}_{ab}(\vn, \uvn) \widehat{\varphi}(\uvn, \vn)\Big) \nn \\
&&- \epsilon_{ab} \psi_b(\vn, \vn) \Big(\varphi(\vn, \uvn) \taub_a(\uvn, \vn) - \widehat{\taub}_a(\vn, \uvn) \widehat{\varphi}(\uvn, \vn)\Big) - 2\epsilon_{ab} \omegab_b(\vn, \vn) \Big(\widehat{\varphib}(\vn, \uvn) \widehat{\psib}_a(\uvn, \vn) \nn \\
&&- \psib_a(\vn, \uvn) \varphib(\uvn, \vn)\Big) + \epsilon_{ab} \chi_{ab}(\vn, \vn) \Big(\widehat{\varphib}(\vn, \uvn) \widehat{\rhob}(\uvn, \vn) - \rhob(\vn, \uvn) \varphib(\uvn, \vn)\Big) \nn \\
&&- \hf \chi_{ab}(\vn, \vn)\Big(\phib(\vn, \uvn) \kappab_{ab}(\uvn, \vn) - \widehat{\kappab}_{ab}(\vn, \uvn)\widehat{\phib}(\uvn, \vn)\Big) - \omegab_a(\vn, \vn)\Big(\phib(\vn, \uvn) \taub_a(\uvn, \vn) \nn \\
&&- \widehat{\taub}_a(\vn, \uvn) \widehat{\phib}(\uvn, \vn)\Big) - \frac{3}{2} \sigmab_{ab}(\vn, \vn)\Big(\widehat{\phi}(\vn, \uvn) \widehat{\xib}_{ab}(\uvn, \vn) - \xib_{ab}(\vn, \uvn)\phi(\uvn, \vn)\Big) \nn \\
&&-2 \Big(\pi(\vn, \vn) \varphi(\vn, \uvn) - \varphi(\vn, \uvn) \widehat{\pi}(\uvn, \uvn)\Big) \Big(\widehat{\pib}(\uvn, \uvn) \varphib(\uvn, \vn) - \varphib(\uvn, \vn) \pib(\vn, \vn)\Big) \nn \\
&&- 2\Big(\pib(\vn, \vn) \phib(\vn, \uvn) - \phib(\vn, \uvn) \widehat{\pib}(\uvn, \uvn)\Big)\Big(\widehat{\pi}(\uvn, \uvn) \phi(\uvn, \vn) - \phi(\uvn, \vn) \pi(\vn, \vn)\Big) \nn \\
&&- 2\Big(\widehat{\varphib}(\vn, \uvn) \widehat{\phib}(\uvn, \vn) - \phib(\vn, \uvn) \varphib(\uvn, \vn)\Big)\Big(\varphi(\vn, \uvn) \phi(\uvn, \vn) - \widehat{\phi}(\vn, \uvn) \widehat{\varphi}(\uvn, \vn)\Big) \nn \\
&&+ \hf \Big(\widehat{\varphib}(\vn, \uvn) \widehat{\varphi}(\uvn, \vn) - \varphi(\vn, \uvn) \varphib(\uvn, \vn) + \phib(\vn, \uvn) \phi(\uvn, \vn) - \widehat{\phi}(\vn, \uvn) \widehat{\phib}(\uvn, \vn)\Big)^2 \nn \\
&&+ [\pib, \pi](\vn, \vn) \Big(\widehat{\varphib}(\vn, \uvn) \widehat{\varphi}(\uvn, \vn) - \varphi(\vn, \uvn) \varphib(\uvn, \vn) + \phib(\vn, \uvn) \phi(\uvn, \vn) \nn \\
&&- \widehat{\phi}(\vn, \uvn) \widehat{\phib}(\uvn, \vn)\Big) \Big\},
\eea

and 

\bea
S^{\rm matter}_{(\Boxb, \Box)} &=& \frac{1}{g_2^2} \sum_{\uvn} \Tr \Big\{2 \varphib(\uvn, \vn) \cDb^{(-)}_a\cD^{(+)}_a \varphi(\vn, \uvn) + 2 \widehat{\phib}(\uvn, \vn) \cDb^{(-)}_a\cD^{(+)}_a \widehat{\phi}(\vn, \uvn) \nn \\
&&+ \Big(\widehat{\cDb}^{(-)}_a \widehat{\cU}_a(\uvn)\Big) \Big(\varphib(\uvn, \vn) \varphi(\vn, \uvn) - \widehat{\varphi}(\uvn, \vn) \widehat{\varphib}(\vn, \uvn) + \widehat{\phib}(\uvn, \vn) \widehat{\phi}(\vn, \uvn) \nn \\
&&- \phi(\uvn, \vn) \phib(\vn, \uvn)\Big) + \widehat{\etab}(\uvn, \vn) \cD^{(-)}_a \widehat{\psib}_a(\vn, \uvn) - \rhob(\uvn, \vn) \cDb^{(-)}_a \taub_a(\vn, \uvn) \nn \\
&&- \kappab_{ab}(\uvn, \vn) \cDb^{(+)}_a \psib_b(\vn, \uvn) - 2 \widehat{\xib}_{ab}(\uvn, \vn) \cD^{(+)}_a \widehat{\taub}_b(\vn, \uvn) \nn \\
&&+ \widehat{\rhob}(\uvn, \vn) \Big(\pi(\vn, \vn) \widehat{\etab}(\vn, \uvn) - \widehat{\etab}(\vn, \uvn) \widehat{\pi}(\uvn, \uvn)\Big) + \taub_a(\uvn, \vn) \Big(\pib(\vn, \vn) \psib_a(\vn, \uvn) \nn \\
&&- \psib_a(\vn, \uvn) \widehat{\pib}(\uvn, \uvn)\Big) + \hf \widehat{\xib}_{ab}(\uvn, \vn) \Big(\pi(\vn, \vn) \widehat{\kappab}_{ab}(\vn, \uvn) - \widehat{\kappab}_{ab}(\vn, \uvn) \widehat{\pi}(\uvn, \uvn)\Big) \nn \\
&&- \widehat{\eta}(\uvn, \uvn) \Big(\widehat{\phib}(\uvn, \vn) \widehat{\etab}(\vn, \uvn) - \etab(\uvn, \vn) \phib(\vn, \uvn)\Big) - 3\epsilon_{ab} \widehat{\eta}(\uvn, \uvn) \Big(\varphib(\uvn, \vn) \xib_{ab}(\vn, \uvn) \nn \\
&&- \widehat{\xib}_{ab}(\uvn, \vn) \widehat{\varphib}(\vn, \uvn)\Big) + \frac{3}{2} \epsilon_{ab} \widehat{\sigmab}_{ab}(\uvn, \uvn) \Big(\widehat{\varphi}(\uvn, \vn) \widehat{\etab}(\vn, \uvn) - \etab(\uvn, \vn) \varphi(\vn, \uvn)\Big) \nn \\
&&- \widehat{\psi}_a(\uvn, \uvn) \Big(\phi(\uvn, \vn) \psib_a(\vn, \uvn) - \widehat{\psib}_a(\uvn, \vn) \widehat{\phi}(\vn, \uvn)\Big) - \widehat{\lambdab}(\uvn, \uvn) \Big(\phi(\uvn, \vn) \rhob(\vn, \uvn) \nn \\
&&- \widehat{\rhob}(\uvn, \vn) \widehat{\phi}(\vn, \uvn)\Big) - \hf \epsilon_{ab} \widehat{\lambdab}(\uvn, \uvn) \Big(\widehat{\varphi}(\uvn, \vn) \widehat{\kappab}_{ab}(\vn, \uvn) - \kappab_{ab}(\uvn, \vn) \varphi(\vn, \uvn)\Big) \nn \\
&&- \epsilon_{ab} \widehat{\psi}_b(\uvn, \uvn) \Big(\widehat{\varphi}(\uvn, \vn) \widehat{\taub}_a(\vn, \uvn) - \taub_a(\uvn, \vn) \varphi(\vn, \uvn)\Big) - 2\epsilon_{ab} \widehat{\omegab}_b(\uvn, \uvn) \Big(\varphib(\uvn, \vn) \psib_a(\vn, \uvn) \nn \\
&&- \widehat{\psib}_a(\uvn, \vn) \widehat{\varphib}(\vn, \uvn)\Big) + \epsilon_{ab} \widehat{\chi}_{ab}(\uvn, \uvn) \Big(\varphib(\uvn, \vn) \rhob(\vn, \uvn) - \widehat{\rhob}(\uvn, \vn) \widehat{\varphib}(\vn, \uvn)\Big) \nn \\
&&- \hf \widehat{\chi}_{ab}(\uvn, \uvn)\Big(\widehat{\phib}(\uvn, \vn) \widehat{\kappab}_{ab}(\vn, \uvn) - \kappab_{ab}(\uvn, \vn) \phib(\vn, \uvn)\Big) - \widehat{\omegab}_a(\uvn, \uvn) \Big(\widehat{\phib}(\uvn, \vn) \widehat{\taub}_a(\vn, \uvn) \nn \\
&&- \taub_a(\uvn, \vn) \phib(\vn, \uvn)\Big) - \frac{3}{2} \widehat{\sigmab}_{ab}(\uvn, \uvn) \Big(\phi(\uvn, \vn) \xib_{ab}(\vn, \uvn) - \widehat{\xib}_{ab}(\uvn, \vn) \widehat{\phi}(\vn, \uvn)\Big) \nn \\
&&-2 \Big(\widehat{\pi}(\uvn, \uvn) \widehat{\varphi}(\uvn, \vn) - \widehat{\varphi}(\uvn, \vn) \pi(\vn, \vn)\Big) \Big(\pib(\vn, \vn) \widehat{\varphib}(\vn, \uvn) - \widehat{\varphib}(\vn, \uvn) \widehat{\pib}(\uvn, \uvn)\Big) \nn \\
&&- 2\Big(\widehat{\pib}(\uvn, \uvn) \widehat{\phib}(\uvn, \vn) - \widehat{\phib}(\uvn, \vn) \pib(\vn, \vn)\Big)\Big(\pi(\vn, \vn) \widehat{\phi}(\vn, \uvn) - \widehat{\phi}(\vn, \uvn) \widehat{\pi}(\uvn, \uvn)\Big) \nn \\
&&- 2\Big(\varphib(\uvn, \vn) \phib(\vn, \uvn) - \widehat{\phib}(\uvn, \vn) \widehat{\varphib}(\vn, \uvn)\Big)\Big(\widehat{\varphi}(\uvn, \vn) \widehat{\phi}(\vn, \uvn) - \phi(\uvn, \vn) \varphi(\vn, \uvn)\Big) \nn \\
&&+ \hf \Big(\varphib(\uvn, \vn) \varphi(\vn, \uvn) - \widehat{\varphi}(\uvn, \vn) \widehat{\varphib}(\vn, \uvn) + \widehat{\phib}(\uvn, \vn) \widehat{\phi}(\vn, \uvn) - \phi(\uvn, \vn) \phib(\vn, \uvn) \Big)^2 \nn \\
&&+ [\widehat{\pib}, \widehat{\pi}](\uvn, \uvn) \Big(\varphib(\uvn, \vn) \varphi(\vn, \uvn) - \widehat{\varphi}(\uvn, \vn) \widehat{\varphib}(\vn, \uvn) + \widehat{\phib}(\uvn, \vn) \widehat{\phi}(\vn, \uvn) \nn \\
&&- \phi(\uvn, \vn) \phib(\vn, \uvn) \Big) \Big\}.
\eea

\begin{figure}
\begin{center}
\includegraphics[width=0.6\textwidth]{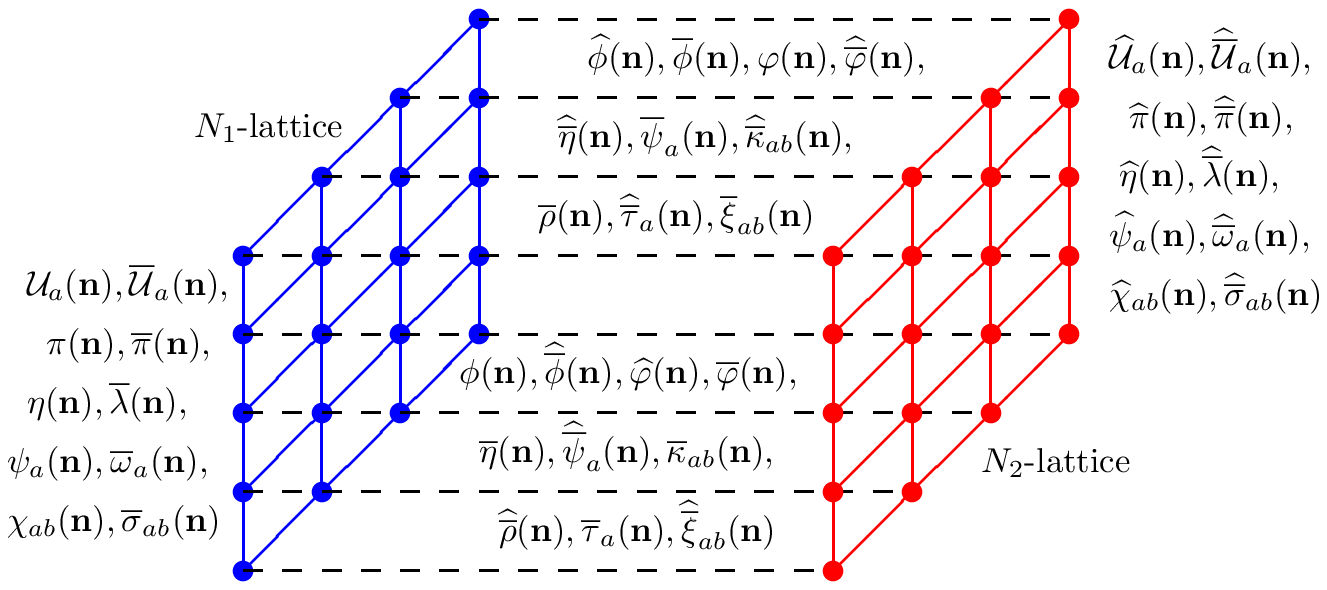}
\end{center}
\caption{\label{fig:2dq8_quiver_latt}Schematic sketch of the lattice construction of two-dimensional $Q=8$ quiver gauge theory. The lattice variables $\{\cU_a(\vn)$, $\cUb_a(\vn)$, $\pi(\vn)$, $\pib(\vn)$, $\eta(\vn)$, $\psi_a(\vn)$, $\chi_{ab}(\vn)$, $\lambdab(\vn)$, $\omegab_a(\vn)$, $\sigmab_{ab}(\vn)\}$ live on the two-dimensional $N_1$-lattice spacetime and $\{\widehat{\cU}_a(\vn)$, $\widehat{\cUb}_a(\vn)$, $\widehat{\pi}(\vn)$, $\widehat{\pib}(\vn)$, $\widehat{\eta}(\vn)$, $\widehat{\psi}_a(\vn)$, $\widehat{\chi}_{ab}(\vn)$, $\widehat{\lambdab}(\vn)$, $\widehat{\omegab}_a(\vn)$, $\widehat{\sigmab}_{ab}(\vn)\}$  live on the two-dimensional $N_2$-lattice spacetime. The matter fields $\{\phi(\vn)$, $\widehat{\phib}(\vn)$, $\widehat{\varphi}(\vn)$, $\varphib(\vn)$, $\etab(\vn)$, $\widehat{\psib}_a(\vn)$, $\kappab_{ab}(\vn)$, $\widehat{\rhob}_a(\vn)$, $\taub_a(\vn)$, $\widehat{\xib}_{ab}(\vn)\}$ and  $\{\widehat{\phi}(\vn)$, $\phib(\vn)$, $\varphi(\vn)$, $\widehat{\varphib}(\vn)$, $\widehat{\etab}(\vn)$, $\psib_a(\vn)$, $\widehat{\kappab}_{ab}(\vn)$, $\rhob_a(\vn)$, $\widehat{\taub}_a(\vn)$, $\xib_{ab}(\vn)\}$ live on the links connecting the two lattice spactimes.}
\end{figure}

In figure \ref{fig:2dq8_quiver_latt} we provide a schematic sketch of the two-dimensional $Q=8$ lattice quiver gauge theory with the placement of adjoint and bi-fundamental fields. 

\subsection{Two-dimensional $Q=8$ lattice circular quiver gauge theory}
\label{sec:2d_q8_lattice_circular_quiver}

The construction of two-dimensional $Q=8$ lattice circular quiver gauge theory with gauge group $U(N_1) \times U(N_2) \times \cdots \times U(N_k)$ is similar to the construction mentioned in the previous section. In this case there are $k$ number of two-dimensional lattice spacetimes, which we label as $N_1$-, $N_2$-, $\cdots$, $N_k$-lattice and they correspond to each node of the circular quiver. The $N_{k+1}$ lattice spacetime is identified with the $N_1$-lattice spacetime. The unit cell is again a square lattice and the twisted fermions are distributed on the unit cell in multiplets of two. Focusing on the three consecutive lattice spacetimes labeled as $i-1$, $i$ and $i+1$, we have the following placement for the adjoint fields. On the node $i$ of the lattice quiver theory we place the adjoint fields $\{\cA_a(\vn)$, $\cAb_a(\vn)$, $\pi(\vn)$, $\pib(\vn)$, $\eta(\vn)$, $\psi_a(\vn)$, $\chi_{ab}(\vn)$, $\lambdab(\vn)$, $\omegab_a(\vn)$, $\sigmab_{ab}(\vn)\}$. On the node $i+1$ we place the adjoint fields $\{\widehat{\cA}_a(\vn)$, $\widehat{\cAb}_a(\vn)$, $\widehat{\pi}(\vn)$, $\widehat{\pib}(\vn)$, $\widehat{\eta}(\vn)$, $\widehat{\psi}_a(\vn)$, $\widehat{\chi}_{ab}(\vn)$, $\widehat{\lambdab}(\vn)$, $\widehat{\omegab}_a(\vn)$, $\widehat{\sigmab}_{ab}(\vn)\}$. On the node $i-1$ we place the following set of adjoint fields $\{\dot{\cA}_a(\vn)$, $\dot{\cAb}_a(\vn)$, $\dot{\pi}(\vn)$, $\dot{\pib}(\vn)$, $\dot{\eta}(\vn)$, $\dot{\psi}_a(\vn)$, $\dot{\chi}_{ab}(\vn)$, $\dot{\lambdab}(\vn)$, $\dot{\omegab}_a(\vn)$, $\dot{\sigmab}_{ab}(\vn)\}$. The bi-fundamental fields connecting the adjacent lattice spacetimes are distributed in the following way. For the nodes $i$ and $i+1$ we have the bi-fundamental fields $\{\phi(\vn)$, $\widehat{\phib}(\vn)$, $\etab(\vn)$, $\widehat{\psib}_a(\vn)$, $\kappab_{ab}(\vn)\}$ transforming in the representation $(\Boxb, \Box)$ while the set of fields $\{\widehat{\phi}(\vn)$, $\phib(\vn)$, $\widehat{\etab}(\vn)$, $\psib_a(\vn)$, $\widehat{\kappab}_{ab}(\vn)\}$ transforming in the representation $(\Box, \Boxb)$. The bi-fundamental fields connecting the the lattice spacetimes $i$ and $i-1$ are distributed in the following way. The set of bi-fundamentals $\{\varphi(\vn)$, $\dot{\varphib}(\vn)$, $\rhob(\vn)$, $\dot{\taub}_a(\vn)$, $\xib_{ab}(\vn)\}$ transforms as $(\Boxb, \Box)$ while the set of fields $\{\dot{\varphi}(\vn)$, $\varphib(\vn)$, $\dot{\rhob}(\vn)$, $\taub_a(\vn)$, $\dot{\xib}_{ab}(\vn)\}$ transforms as $(\Box, \Boxb)$.

\subsection{Three-dimensional $Q=8$ lattice quiver gauge theory}
\label{sec:3d_q8_lattice_quiver}

\begin{figure}
\begin{center}
\includegraphics[width=0.45\textwidth]{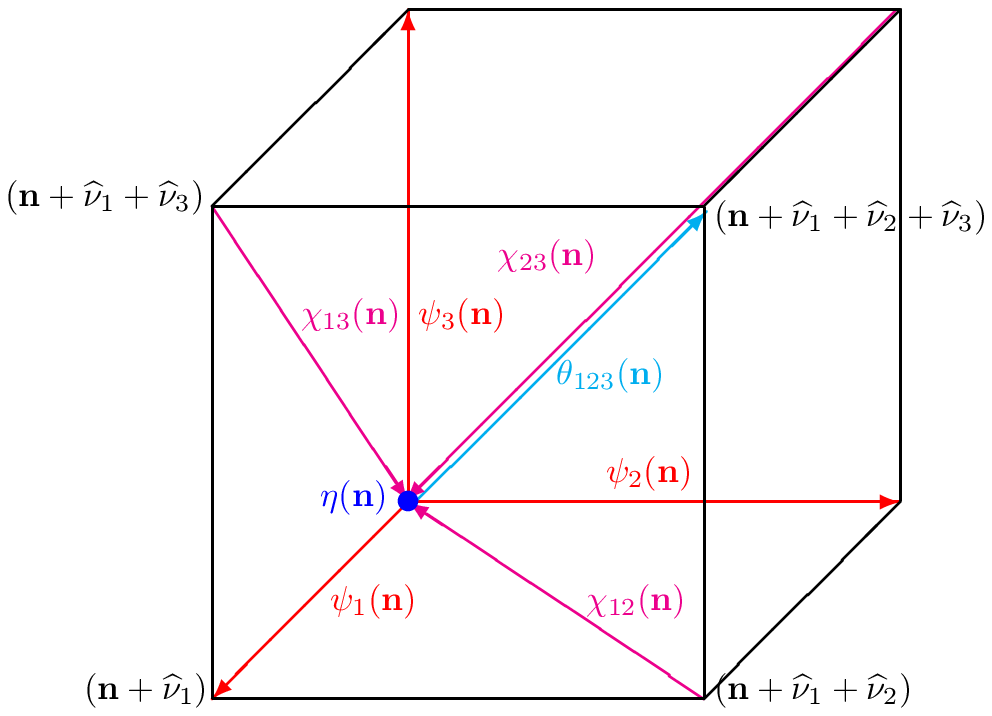}
\end{center}
\caption{\label{fig:3dq8latt}The unit cell of three-dimensional $Q=8$ lattice SYM with orientation assignments for the twisted fermions. The complexified bosons $\cU_m$ follow the same orientations and link assignments as that of their superpartners $\psi_m$.}
\end{figure}

The three-dimensional $Q=8$ lattice quiver gauge theory with gauge group $U(N_1) \times U(N_2)$ can be constructed in a way similar to the constructions mentioned above. The lattice theory contains two three-dimensional lattice spacetimes ($N_1$- and $N_2$-lattice) corresponding to each node of the quiver. The unit cell is a cubic lattice. The orientations and placements of the twisted fermions $\eta(\vn)$, $\psi_m(\vn)$, $\chi_{mn}(\vn)$ and $\theta_{mnr}(\vn)$ are illustrated in figure \ref{fig:3dq8latt}. They all transform in the adjoint representation of the gauge group $U(N_1)$. The complexified Wilson links $\cU_m(\vn)$ have the same orientations and placements as that of their superpartners $\psi_m(\vn)$. The fields $\cUb_m(\vn)$ have the same placements but opposite orientations compared to $\cU_m(\vn)$. The placements and orientations of the twisted fields on the lattice respect gauge-invariance and scalar supersymmetry of the lattice theory. The set of adjoint fields that are decorated with hats live on the $N_2$-lattice spacetime and transform as adjoints under the gauge group $U(N_2)$. 

The action of the three-dimensional $Q=8$ lattice quiver gauge theory has the following form
\beq
S = S^{\rm SYM}_{({\bf adj}, {\bf 1})} + S^{\rm SYM}_{({\bf 1}, {\bf adj})} + S^{\rm matter}_{(\Box, \Boxb)} + S^{\rm matter}_{(\Boxb, \Box)},
\eeq
where

\bea
S^{\rm SYM}_{({\bf adj}, {\bf 1})} &=& \frac{1}{g_3^2}\sum_{\vn} \Tr \Big\{\Big(\cUb_n(\vn + \hatbnu_m)\cUb_m(\vn) - \cUb_m(\vn + \hatbnu_n)\cUb_n(\vn)\Big) \nn \\
&&~~~~\times \Big(\cU_{m}(\vn)\cU_{n}(\vn + \hatbnu_m) - \cU_n(\vn)\cU_m(\vn + \hatbnu_n)\Big) \nn \\
&&~~~~+ \hf \Big(\cU_m(\vn)\cUb_m(\vn) - \cUb_m(\vn - \hatbnu_m)\cU_m(\vn - \hatbnu_m)\Big)^2 \nn \\
&&~~~~+\hf(\delta_{mq}\delta_{nr} - \delta_{mr}\delta_{nq}) \chi_{mn}(\vn) \Big(\cU_q(\vn)\psi_r(\vn + \hatbnu_q) - \psi_r(\vn)\cU_q(\vn + \hatbnu_r)\Big) \nn \\
&&~~~~+ \eta(\vn)\Big(\psi_m(\vn)\cUb_m(\vn) - \cUb_m(\vn - \hatbnu_m)\psi_m(\vn - \hatbnu_m)\Big) \nn \\
&&~~~~+\frac{1}{3}(\delta_{mr}\delta_{ne}\delta_{qf} + \delta_{qr}\delta_{me}\delta_{nf} + \delta_{nr}\delta_{qe}\delta_{mf}) \nn \\
&&~~~~\times \theta_{ref}(\vn)\Big(\chi_{re}(\vn+\hatbnu_f)\cUb_f(\vn) - \cUb_f(\vn+\hatbnu_r+\hatbnu_e)\chi_{re}(\vn)\Big)\Big\},
\eea

\bea
S^{\rm SYM}_{({\bf 1}, {\bf adj})} &=& \frac{1}{g_3^2}\sum_{\uvn} \Tr \Big\{\Big(\widehat{\cUb}_n(\uvn + \uhatbnu_m)\widehat{\cUb}_m(\uvn) - \widehat{\cUb}_m(\uvn + \uhatbnu_n)\widehat{\cUb}_n(\uvn)\Big) \nn \\
&&~~~~\times \Big(\widehat{\cU}_{m}(\uvn)\widehat{\cU}_{n}(\uvn + \uhatbnu_m) - \widehat{\cU}_n(\uvn)\widehat{\cU}_m(\uvn + \uhatbnu_n)\Big) \nn \\
&&~~~~+ \hf \Big(\widehat{\cU}_m(\uvn)\widehat{\cUb}_m(\uvn) - \widehat{\cUb}_m(\uvn - \uhatbnu_m)\widehat{\cU}_m(\uvn - \uhatbnu_m)\Big)^2 \nn \\
&&~~~~+ \hf(\delta_{mq}\delta_{nr} - \delta_{mr}\delta_{nq}) \widehat{\chi}_{mn}(\uvn) \Big(\widehat{\cU}_q(\uvn)\widehat{\psi}_r(\uvn + \uhatbnu_q) - \widehat{\psi}_r(\uvn)\widehat{\cU}_q(\uvn + \uhatbnu_r)\Big) \nn \\
&&~~~~+ \widehat{\eta}(\uvn)\Big(\widehat{\psi}_m(\uvn)\widehat{\cUb}_m(\uvn) - \widehat{\cUb}_m(\uvn - \uhatbnu_m)\widehat{\psi}_m(\uvn - \uhatbnu_m)\Big) \nn \\
&&~~~~+\frac{1}{3}(\delta_{mr}\delta_{ne}\delta_{qf} + \delta_{qr}\delta_{me}\delta_{nf} + \delta_{nr}\delta_{qe}\delta_{mf}) \nn \\
&&~~~~\times \widehat{\theta}_{ref}(\uvn)\Big(\widehat{\chi}_{re}(\uvn+\uhatbnu_f)\widehat{\cUb}_f(\uvn) - \widehat{\cUb}_f(\uvn+\uhatbnu_r+\uhatbnu_e)\widehat{\chi}_{re}(\uvn)\Big)\Big\},
\eea

\bea
S^{\rm matter}_{(\Box, \Boxb)} &=& \frac{1}{g_3^2}\sum_{\vn} \Tr \Big\{2 \widehat{\varphib}(\vn, \uvn) \cDb^{(-)}_m \cD^{(+)}_m \widehat{\varphi}(\uvn, \vn) + 2 \phib(\vn, \uvn) \cDb^{(-)}_m \cD^{(+)}_m \phi(\uvn, \vn) \nn \\
&&~~~~+ \Big(\cDb^{(-)}_m \cU_m(\vn)\Big) \Big(\widehat{\varphib}(\vn, \uvn) \widehat{\varphi}(\uvn, \vn) - \varphi(\vn, \uvn) \varphib(\uvn, \vn) + \phib(\vn, \uvn) \phi(\uvn, \vn) \nn \\
&&~~~~- \widehat{\phi}(\vn, \uvn) \widehat{\phib}(\uvn, \vn) \Big) + \etab(\vn, \uvn) \cD^{(-)}_m \psib_m(\uvn, \vn) - \widehat{\kappab}_{np}(\vn, \uvn)\cDb^{(+)}_p \widehat{\psib}_n(\uvn, \vn) \nn \\
&&~~~~+ \hf \thetab_{mnr}(\vn, \uvn) \cD^{(+)}_m \kappab_{nr}(\uvn, \vn) - \hf \epsilon_{mnr}\psi_r(\vn, \vn)\Big(\varphi(\vn, \uvn) \kappab_{mn}(\uvn, \vn) \nn \\
&&~~~~- \widehat{\kappab}_{mn}(\vn, \uvn) \widehat{\varphi}(\uvn, \vn)\Big) + \epsilon_{mnr}\theta_{mnr}(\vn, \vn) \Big(\varphi(\vn, \uvn) \etab(\uvn, \vn) \nn \\
&&~~~~- \widehat{\etab}(\vn, \uvn) \widehat{\varphi}(\uvn, \vn)\Big) - \psi_m(\vn, \vn) \Big(\widehat{\phi}(\vn, \uvn) \widehat{\psib}_m(\uvn, \vn) - \psib_m(\vn, \uvn) \phi(\uvn, \vn)\Big) \nn \\
&&~~~~- \theta_{mnr}(\vn, \vn) \Big(\widehat{\phi}(\vn, \uvn) \widehat{\thetab}_{mnr}(\uvn, \vn) - \thetab_{mnr}(\vn, \uvn) \phi(\uvn, \vn)\Big) \nn \\
&&~~~~- \epsilon_{mnr} \eta(\vn, \vn) \Big(\widehat{\varphib}(\vn, \uvn) \widehat{\thetab}_{mnr}(\uvn, \vn) - \thetab_{mnr}(\vn, \uvn) \varphib(\uvn, \vn)) \nn \\
&&~~~~- \eta(\vn, \vn) (\phib(\vn, \uvn) \etab(\uvn, \vn) - \widehat{\etab}(\vn, \uvn) \widehat{\phib}(\uvn, \vn)\Big) - \hf \chi_{mn}(\vn, \vn) \Big(\phib(\vn, \uvn)  \kappab_{mn}(\uvn, \vn) \nn \\
&&~~~~- \widehat{\kappab}_{mn}(\vn, \uvn) \widehat{\phib}(\uvn, \vn)\Big) + \epsilon_{mnr} \chi_{mn}(\vn, \vn) \Big(\widehat{\varphib}(\vn, \uvn) \widehat{\psib}_r(\uvn, \vn) - \psib_r(\vn, \uvn) \varphib(\uvn, \vn)\Big) \nn \\
&&~~~~+ \hf \Big(\widehat{\varphib}(\vn, \uvn) \widehat{\varphi}(\uvn, \vn) - \varphi(\vn, \uvn) \varphib(\uvn, \vn) + \phib(\vn, \uvn) \phi(\uvn, \vn) - \widehat{\phi}(\vn, \uvn) \widehat{\phib}(\uvn, \vn) \Big)^2 \nn \\
&&~~~~- 2 (\widehat{\varphib}(\vn, \uvn) \widehat{\phib}(\uvn, \vn) - \phib(\vn, \uvn) \varphib(\uvn, \vn))(\varphi(\vn, \uvn) \phi(\uvn, \vn) - \widehat{\phi}(\vn, \uvn) \widehat{\varphi}(\uvn, \vn)) \Big\},~~~~~~~~~~
\eea

and

\bea
S^{\rm matter}_{(\Boxb, \Box)} &=& \frac{1}{g_3^2}\sum_{\uvn} \Tr \Big\{2 \varphib(\uvn, \vn) \cDb^{(-)}_m \cD^{(+)}_m \varphi(\vn, \uvn) + 2 \widehat{\phib}(\uvn, \vn) \cDb^{(-)}_m \cD^{(-)}_m \widehat{\phi}(\vn, \uvn) \nn \\
&&~~~~+ \Big(\cDb^{(-)}_m \cU_m(\uvn)\Big) \Big(\varphib(\uvn, \vn) \varphi(\vn, \uvn) - \widehat{\varphi}(\uvn, \vn) \widehat{\varphib}(\vn, \uvn) + \widehat{\phib}(\uvn, \vn) \widehat{\phi}(\vn, \uvn)  \nn \\
&&~~~~- \phi(\uvn, \vn) \phib(\vn, \uvn) \Big) + \widehat{\etab}(\uvn, \vn) \cD^{(-)}_m \widehat{\psib}_m(\vn, \uvn) - \kappab_{np}(\uvn, \vn) \cDb^{(+)}_p \psib_n(\vn, \uvn) \nn \\
&&~~~~+ \hf \widehat{\thetab}_{mnr}(\uvn, \vn) \cD^{(+)}_m \widehat{\kappab}_{nr}(\vn, \uvn) - \hf \epsilon_{mnr}\widehat{\psi}_r(\uvn, \uvn) \Big(\widehat{\varphi}(\uvn, \vn) \widehat{\kappab}_{mn}(\vn, \uvn) \nn \\
&&~~~~- \kappab_{mn}(\uvn, \vn) \varphi(\vn, \uvn)\Big) + \epsilon_{mnr} \widehat{\theta}_{mnr} (\uvn, \uvn)\Big(\widehat{\varphi}(\uvn, \vn) \widehat{\etab}(\vn, \uvn) \nn \\
&&~~~~- \etab(\uvn, \vn) \varphi(\vn, \uvn)\Big) - \widehat{\psi}_m(\uvn, \uvn) \Big(\phi(\uvn, \vn) \psib_m(\vn, \uvn) - \widehat{\psib}_m(\uvn, \vn) \widehat{\phi}(\vn, \uvn)\Big) \nn \\
&&~~~~- \widehat{\theta}_{mnr}(\uvn, \uvn) \Big(\phi(\uvn, \vn) \thetab_{mnr}(\vn, \uvn) - \widehat{\thetab}_{mnr}(\uvn, \vn) \widehat{\phi}(\vn, \uvn)\Big) \nn \\
&&~~~~- \epsilon_{mnr} \widehat{\eta}(\uvn, \uvn) \Big(\varphib(\uvn, \vn) \thetab_{mnr}(\vn, \uvn) - \widehat{\thetab}_{mnr}(\uvn, \vn) \widehat{\varphib}(\vn, \uvn)\Big) \nn \\
&&~~~~- \widehat{\eta}(\uvn, \uvn) \Big(\widehat{\phib}(\uvn, \vn) \widehat{\etab}(\vn, \uvn) - \etab(\uvn, \vn) \phib(\vn, \uvn)\Big) - \hf \widehat{\chi}_{mn}(\uvn, \uvn) \Big(\widehat{\phib}(\uvn, \vn)  \widehat{\kappab}_{mn}(\vn, \uvn) \nn \\
&&~~~~- \kappab_{mn}(\uvn, \vn) \phib(\vn, \uvn)\Big) + \epsilon_{mnr} \widehat{\chi}_{mn}(\uvn, \uvn) \Big(\varphib(\uvn, \vn) \psib_r(\vn, \uvn) - \widehat{\psib}_r(\uvn, \vn) \widehat{\varphib}(\vn, \uvn)\Big) \nn \\
&&~~~~+ \hf \Big(\varphib(\uvn, \vn) \varphi(\vn, \uvn) - \widehat{\varphi}(\uvn, \vn) \widehat{\varphib}(\vn, \uvn) + \widehat{\phib}(\uvn, \vn) \widehat{\phi}(\vn, \uvn) - \phi(\uvn, \vn) \phib(\vn, \uvn) \Big)^2 \nn \\
&&~~~~- 2 (\varphib(\uvn, \vn) \phib(\vn, \uvn) - \widehat{\phib}(\uvn, \vn) \widehat{\varphib}(\vn, \uvn))(\widehat{\varphi}(\uvn, \vn) \widehat{\phi}(\vn, \uvn) - \phi(\uvn, \vn) \varphi(\vn, \uvn)) \Big\}.~~~~~~~~~~
\eea

\begin{figure}
\begin{center}
\includegraphics[width=0.6\textwidth]{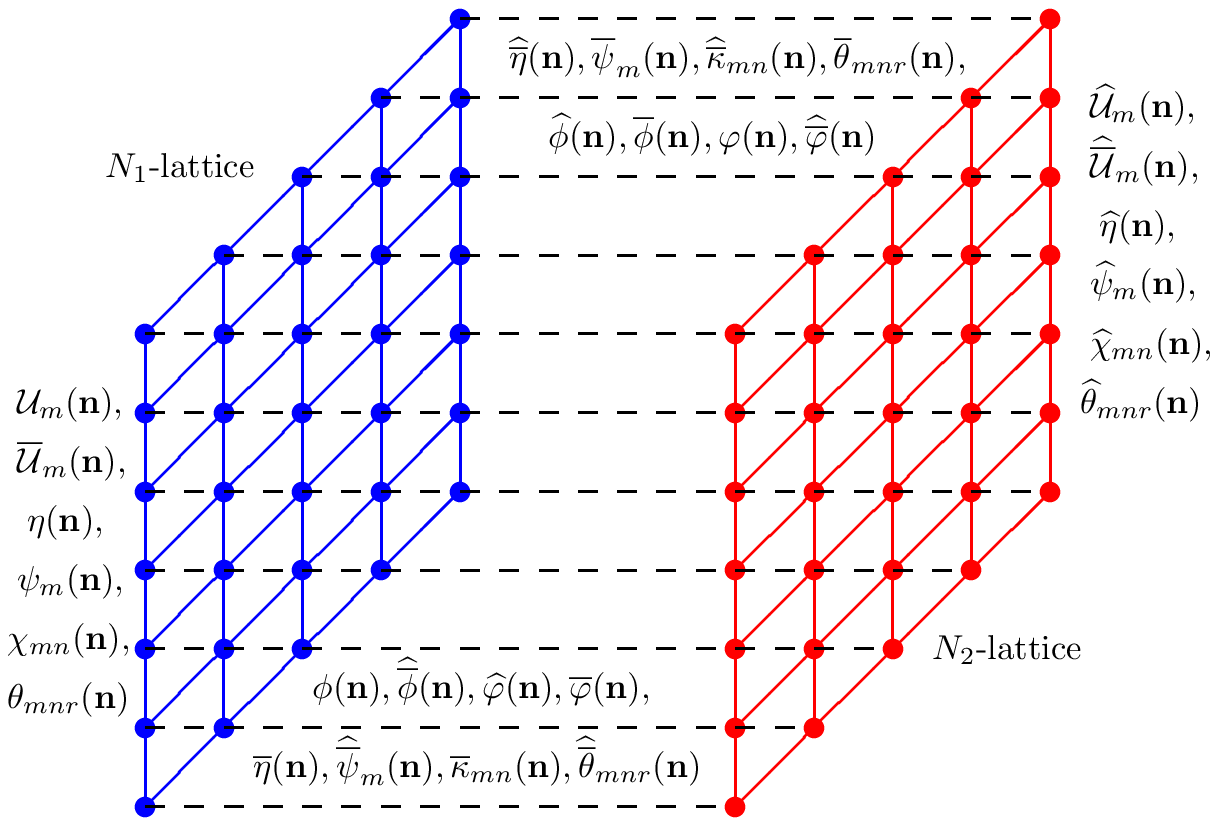}
\end{center}
\caption{\label{fig:3dq8_quiver_latt}Schematic sketch of the lattice construction of three-dimensional $Q=8$ quiver gauge theory. The lattice variables $\{\cU_m(\vn)$, $\cUb_m(\vn)$, $\eta(\vn)$, $\psi_m(\vn)$, $\chi_{mn}(\vn)$, $\theta_{mnr}(\vn)\}$ live on the three-dimensional $N_1$-lattice spacetime and $\{\widehat{\cU}_m(\vn)$, $\widehat{\cUb}_m(\vn)$, $\widehat{\eta}(\vn)$, $\widehat{\psi}_m(\vn)$, $\widehat{\chi}_{mn}(\vn)$, $\widehat{\theta}_{mnr}(\vn)\}$ live on the three-dimensional $N_2$-lattice spacetime. The matter fields $\{\phi(\vn)$, $\widehat{\phib}(\vn)$, $\widehat{\varphi}(\vn)$, $\varphib(\vn)$, $\etab(\vn)$, $\widehat{\psib}_m(\vn)$, $\kappab_{mn}(\vn)$, $\widehat{\thetab}_{mnr}(\vn)\}$ and  $\{\widehat{\phi}(\vn)$, $\phib(\vn)$, $\varphi(\vn)$, $\widehat{\varphib}(\vn)$, $\widehat{\etab}(\vn)$, $\psib_m(\vn)$, $\widehat{\kappab}_{mn}(\vn)$, $\thetab_{mnr}(\vn)\}$ live on the links connecting the two lattice spactimes.}
\end{figure}

In figure \ref{fig:3dq8_quiver_latt} we provide a schematic sketch of the three-dimensional $Q=8$ lattice quiver gauge theory with the placement of adjoint and bi-fundamental fields on the lattice quiver. 

\section{Discussions and prospects}
\label{sec:discussions_prospects}

In this paper, we have detailed the constructions of several classes of twisted supersymmetric quiver gauge theories in the continuum and also on the lattice. These theories live on two and three Euclidean spacetime dimensions and possess four and eight supercharges. The process of twisting allows us to easily transport these theories on to the lattice by preserving a subset of supersymmetries exact at finite lattice spacing. We used the method of geometric discretization to implement these theories on the lattice. The lattice theories constructed this way are gauge-invariant, doubler free and retain at least one supercharge exact on the lattice. In two dimensions we have discussed the constructions of lattice quiver gauge theories possessing four and eight supercharges. We have also constructed two-dimensional eight supercharge lattice quiver gauge theory with circular topology. We have also provided the continuum and lattice constructions of three-dimensional quiver gauge theory possessing eight supercharges. These quiver theories contain adjoint fields living on the nodes and bi-fundamental matter fields linking between the nodes. It is interesting to note that we can also construct lattice gauge theories with fundamental matter fields, from these quiver gauge theories, by retaining one node of the lattice quiver theory and freezing the rest of the nodes. This process of freezing the degrees of freedom is not in conflict with supersymmetry. Such constructions have been carried out for the two dimensional $\cN=(2, 2)$ case in ref. \cite{Matsuura:2008cfa} and the three-dimensional $\cN=4$ case in ref. \cite{Joseph:2013jya}.

The construction of two-dimensional four supercharge lattice quiver gauge theory has the realization as a system of intersecting branes. In the lattice quiver theory there are two lattice spacetimes and the bi-fundamental matter fields are the lattice variables living on links connecting them. The two-dimensional four supercharge lattice quiver gauge theory could be realized as a system of intersecting branes in type IIA string theory \cite{Orlando:2010uu}. There are D2 branes stretching between two NS5 branes. The brane configurations are arranged in such a way that they preserve 4 out of the 32 supercharges of type IIA string theory. Open strings stretching between the D2 branes located between parallel NS5 branes correspond to the adjoint fields in the $\cN=(2, 2)$ sector, while open strings stretching between two separate stacks of D2 branes correspond to bi-fundamental fields of the quiver theory.

The two dimensional $\cN = (2, 2)$ supersymmetric quiver gauge theories can be related to quantum integrable systems, such as spin chains, through the Gauge/Bethe correspondence. One can use the correspondence to map the supersymmetric ground states of the gauge theory directly to the Bethe spectrum of the integrable model \cite{Orlando:2010uu}.

We can also look at the three-dimensional eight supercharge ($\cN = 4$) lattice quiver gauge theory in the context of intersecting branes. Three-dimensional $\cN=4$ quiver gauge theories admit realization as low-energy limit brane configurations of Hanany-Witten type \cite{Hanany:1996ie} in type IIB string theory. The string theory contains D3 branes that are stretched between NS5 and D5 branes such that the fivebranes have two common world-volume directions and one common transverse direction. The D3 brane is wrapped on the two common world-volume directions and the common transverse direction. In the field theory limit and at energy scales below the scale set by the interval between the NS5 branes, the world-volume theories on the D3 branes become three-dimensional $U(N)$ $\cN=4$ SYM gauge theories, giving rise to the desired quiver. Three-dimensional $\cN=4$ gauge theories also play an important role in our understanding of dualities. First examples of three-dimensional mirror symmetry \cite{Intriligator:1996ex, Hanany:1996ie, deBoer:1996mp, deBoer:1996ck, Kapustin:1999ha} were provided by such theories and we hope that the lattice theory constructed here would be useful for non-perturbative investigations related to such dualities.

\acknowledgments

I thank discussions and communications with Simon Catterall, Joel Giedt, David B. Kaplan, So Matsuura and Mithat \"Unsal on various aspects of exact lattice supersymmetry. This work was supported in part by the DFG Sonderforschungsbereich/Transregio SFB/TR9.

\end{document}